\documentclass[useAMS,usenatbib]{mn2e}
\usepackage{longtable,lscape}
\usepackage{amsmath}

\usepackage{graphicx}
\usepackage{aas_macros}
\usepackage{amssymb}

\newcommand{\about}{$\sim\!\!$~}
\newcommand{\kms}{km~s$^{-1}$}
\newcommand{\etal}{et~al.\ }
\newcommand{\msun}{M$_\odot$}

\newcommand{\bvri}{\protect\hbox{$BV\!RI$} }

\newcommand{\bvmax}{\protect\hbox{$\left(B-V\right)_{\rm max}$}}

\mathchardef\mhyphen="2D

\newcommand{\be}{\begin{displaymath}}
\newcommand{\ee}{\end{displaymath}}

\def\lsim{\hbox{\rlap{\raise 0.425ex\hbox{$<$}}\lower 0.65ex\hbox{$\sim$}}}
\def\gsim{\hbox{\rlap{\raise 0.425ex\hbox{$>$}}\lower 0.65ex\hbox{$\sim$}}}

\newcommand{\ion}[2]{#1$\;${\small{#2}}\relax}

\graphicspath{{/Users/jsilv/BSNIP/carbon/plots/}}

\title[Carbon in SN~Ia Spectra]{Carbon Detection in Early-Time 
Optical Spectra of Type~Ia Supernovae} 

\author[Silverman, et~al.]{Jeffrey~M.~Silverman,$^{1}$\thanks{E-mail:
    JSilverman@astro.berkeley.edu} Alexei~V.~Filippenko$^{1}$ \\
$^{1}$Department of Astronomy, University of California, Berkeley, CA 94720-3411, USA \\
}

\begin{document}
\date{Accepted  . Received   ; in original form  }
\pagerange{\pageref{firstpage}--\pageref{lastpage}} \pubyear{2012}
\maketitle
\label{firstpage}

\begin{abstract}
While O is often seen in spectra of Type~Ia supernovae (SNe~Ia) as both
unburned fuel and a product of C burning, C is only occasionally seen at
the earliest times, and it represents the most direct way of 
investigating primordial white dwarf material and its relation to SN~Ia 
explosion scenarios and mechanisms. In this paper, we search for C
absorption features in 188 optical spectra of 144 low-redshift ($z <
0.1$) SNe~Ia with ages $\la$3.6~d after maximum brightness. These data
were obtained as part of the Berkeley SN~Ia Program (BSNIP; Silverman
\etal 2012) and represent the largest set of SNe~Ia in 
which C has ever been searched. We find that \about11~per~cent of the
SNe studied show definite C absorption features while
\about25~per~cent show some evidence for \ion{C}{II} in their
spectra. Also, if 
one obtains a spectrum at $t \la -5$~d, then there is a better than
30~per~cent chance of detecting a distinct absorption feature from
\ion{C}{II}. SNe~Ia that show C are found to resemble those without C
in many respects, but objects with C tend to have bluer optical 
colours than those without C. The typical expansion velocity of the
\ion{C}{II} $\lambda$6580 feature is measured to be
12,000--13,000~\kms, and the ratio of the \ion{C}{II} $\lambda$6580 to
\ion{Si}{II} $\lambda$6355 velocities is remarkably constant with time
and among different objects with a median value of \about1.05. While
the pseudo-equivalent widths (pEWs) of the \ion{C}{II} $\lambda$6580 and
\ion{C}{II} $\lambda$7234 features are found mostly to decrease with
time, we see evidence of a significant increase in pEW between \about12 
and 11~d before maximum brightness, which is actually predicted by some
theoretical models. The range of pEWs measured from the BSNIP
data implies a range of C mass in SN~Ia ejecta of about
$\left(2\textrm{--}30\right) \times 10^{-3}$~\msun. 
\end{abstract}



\section{Introduction}\label{s:intro}

It is thought that thermonuclear explosions of C/O white dwarfs (WDs)
give rise to Type~Ia supernovae (SNe~Ia; e.g., \citealt{Hoyle60,
  Colgate69, Nomoto84}; see \citealt{Hillebrandt00} for a
review). However, after decades of observations and theoretical work,
the details of SN~Ia progenitors and explosion mechanisms are still
missing. Despite this, SNe~Ia have been used in the recent past to
discover the accelerating
expansion of the Universe \citep{Riess98:lambda,Perlmutter99}, as well as
to measure cosmological parameters \citep[e.g.,][]{Astier06,
  Riess07,Wood-Vasey07, Hicken09:cosmo, Kessler09,
  Amanullah10,Suzuki12}.

As the explosion proceeds, pristine C and O from the progenitor WD may
get mixed throughout various layers of the 
ejecta. However, the amount and exact location of this unburned
material varies widely among published models
\citep[e.g.,][]{Hoflich02,Gamezo03,Ropke07,Kasen09}. Thus,
determinations of the quantity of these elements after explosion,
their spatial distribution in the ejecta, and other measurements of
this primordial material can help constrain possible explosion
mechanisms of SNe~Ia. 

Oxygen is found in SNe~Ia as both unburned fuel from the progenitor WD
and burned ash as a product of C burning. Therefore, it is
difficult to associate O detections, which are common in SNe~Ia
\citep[e.g.,][]{Filippenko97}, with primordial material. This leaves
us with C as the most direct link to matter from the
pre-explosion WD. Observations during the first 2--3 weeks after
explosion probe the outermost layers of the ejecta, which is where the
unburned material is most likely to reside. At the temperatures and
densities observed in SNe~Ia at these epochs, singly ionised states of
C are expected to be the dominant species 
\citep[e.g.,][]{Tanaka08}. Neutral C could appear (mostly in the
near-infrared) at significantly lower temperatures, and doubly ionised C
would require much higher temperatures \citep{Marion06}. Thus, the
best chance of detecting unburned material is to look for \ion{C}{II}
features before and near $B$-band maximum brightness.

Until recently, C detections in SNe~Ia were only noticed in a small
handful of objects. Most of the SNe~Ia that show obvious \ion{C}{II}
absorption features are extremely luminous objects with slowly
evolving light curves and exceptionally low expansion velocities. These
objects are thought to arise from super-Chandrasekhar-mass WDs
\citep{Howell06,Yamanaka09,Scalzo10,Silverman11,Taubenberger11}, and
thus the detection of unburned C in their spectra is likely related to
the relatively rare explosion mechanism that produces these
objects. In addition, there are a few instances of relatively normal
SNe~Ia \citep[i.e., ones that obey the relation between light-curve 
decline rate and luminosity at peak brightness, known as the ``Phillips
relation'';][]{Phillips93} that also show strong \ion{C}{II}
absorption \citep[e.g.,][]{Patat96,Garavini05}. On the other hand, it
has usually been found that C does not appear at all in early-time
optical spectra of SNe~Ia, or it is weak or extremely blended
\citep[e.g.,][]{Mazzali01,Branch03,Stanishev07,Thomas07}. 

Recently, however, C detections in SNe~Ia at early times have become
more common thanks to the amassing of many more spectra of SNe~Ia at
very early epochs, higher signal-to-noise ratio (S/N) data, and
astronomers being more meticulous in their search for this elusive
unburned material. \citet{Parrent11} present new spectra of 3 SNe~Ia
that show distinct C absorptions and analyze those alongside 65 other
objects from the literature. They find that \ion{C}{II} features are
detected more often than previously thought and estimate that
30~per~cent of all SNe~Ia may show evidence of unburned C in their
pre-maximum spectra. This work is explained by \citet{Thomas11}, who
discuss observations and analyses of 5 more objects that show
\ion{C}{II} features. They conclude that the C is likely
distributed spherically symmetrically and that
$22^{+10}_{-6}$~per~cent of SNe~Ia show C absorption features at
epochs near 5~d before maximum brightness. Finally,
\citet{Folatelli11} use data from the Carnegie Supernova Project to
determine that at least 30~per~cent of objects show \ion{C}{II}
absorption and that the mass of C in the ejecta is consistent with
$10^{-3}$--$10^{-2}$~\msun. They also find evidence that SNe~Ia with C
tend to have bluer colours and lower luminosities at maximum light.

In this work, we search for possible C signatures in low-redshift ($z
< 0.1$) optical spectra of SNe~Ia obtained as part of the Berkeley
SN~Ia Program (BSNIP). The data are presented in BSNIP~I
\citep{Silverman12:BSNIPI}, and we utilise the spectral feature
measurement tools described in BSNIP~II \citep*{Silverman12:BSNIPII}. 
With such a large, self-consistent
dataset, we are able to accurately explore the incidence rate of C in SN~Ia
spectra and how it varies as a function of observed
epoch. Furthermore, we can quantify the amount and location in the
ejecta of the unburned C by measuring 
pseudo-equivalent widths (pEWs) and expansion velocities,
respectively.

The spectral and photometric data used herein are summarised in
Section~\ref{s:data}, and our methods for determining the presence or
absence of C and (for SNe~Ia with definite C detection) measuring
\ion{C}{II} spectral features is described in Section~\ref{s:procedure}. 
Section~~\ref{s:analysis} presents the rate of C detection, 
the differences and similarities between SNe~Ia with and without C, 
and a discussion of the \ion{C}{II} spectral feature measurements.
We present our conclusions in Section~\ref{s:conclusions}.


\section{Dataset}\label{s:data}

The SN~Ia spectral data investigated in the current study are a subset of those
used in BSNIP~II and originally published in BSNIP~I. The majority of
the spectra were obtained using the Shane 3~m telescope at Lick
Observatory with the Kast double spectrograph \citep{Miller93}, and
the typical wavelength coverage is 3300--10,400~\AA\ with resolutions
of \about11 and \about6~\AA\ on the red and blue sides (crossover
wavelength \about5500~\AA),  respectively. For more information
regarding the observations and data reduction, see BSNIP~I.

In BSNIP~II, we ignored {\it a priori} the extremely peculiar
SN~2000cx \citep[e.g.,][]{Li01:00cx}, SN~2002cx
\citep[e.g.,][]{Li03:02cx,Jha06:02cx}, SN~2005hk
\citep[e.g.,][]{Chornock06,Phillips07}, and SN~2008ha
\citep[e.g.,][]{Foley09:08ha,Valenti09}. This was mainly due to the
fact that they are so spectroscopically distinct from the bulk of the
SN~Ia population that their spectral features are difficult to measure
in the same way as for the other objects. It should be noted that
\citet{Parrent11} find that all of these objects show evidence for
unburned C. However, in this work we will only concentrate on SNe~Ia
that follow the Phillips relation, and thus can be used as
cosmological distance indicators. This means that we also remove all
super-Chandrasekhar-mass SNe~Ia from our sample, even though they show
strong absorption from \ion{C}{II} (as mentioned above). 

BSNIP~II contains 432 spectra of 261 SNe~Ia with ages younger than
20~d (rest frame) past maximum brightness. We began the current study
by inspecting all 206 spectra (of 156 objects) younger than 5~d past
maximum for possible C signatures. 
The oldest spectrum to show evidence of a \ion{C}{II} feature was
obtained \about3.6~d after maximum. Hence, for the rest of this study,
we only consider spectra younger than this epoch. This yields a sample
of 188 spectra of 144 SNe~Ia, which is the largest set of SNe~Ia that
has ever been inspected for C features. A summary of these objects,
their ``Carbon Classification'' (see Section~\ref{s:procedure}), and
their spectral classifications based on various classification schemes
can be found in Table~\ref{t:objects}. For comparison,
\citet{Parrent11} investigated 58 objects\footnote{The dataset only includes
  SNe~Ia that follow the Phillips relation.} younger than 1~d past 
maximum brightness, \citet{Thomas11} used 124 objects at epochs before
2.5~d past maximum, and the study by \citet{Folatelli11} utilised 51
SNe~Ia with spectra before maximum.

The spectral ages of the BSNIP data referred to throughout this work
are calculated using the redshift and Julian Date of $B$-band maximum
brightness presented
in Table~1 of BSNIP~I. Furthermore, photometric parameters (such as 
light-curve width and colour information) used in the present study
can be found in Ganeshalingam \etal (in preparation). 

\setlength{\tabcolsep}{0.05in}
\onecolumn
\begin{center}
\begin{longtable}{lccccc|lccccc}
\caption{Summary of SNe~Ia in the Sample}\label{t:objects} \\[-2ex]
\hline \hline
SN Name & \ion{C}{II} & SNID & Benetti & Branch & Wang & SN Name & \ion{C}{II} & SNID & Benetti & Branch & Wang \\
  & Type$^\textrm{a}$ & (Sub)Type$^\textrm{b}$ & Type$^\textrm{c}$ & Type$^\textrm{d}$ & Type$^\textrm{e}$ &  & Type$^\textrm{a}$ & (Sub)Type$^\textrm{b}$ & Type$^\textrm{c}$ & Type$^\textrm{d}$ & Type$^\textrm{e}$ \\
\hline
\endfirsthead
\multicolumn{12}{c}{{\tablename} \thetable{} --- Continued} \\
\hline \hline
SN Name & \ion{C}{II} & SNID & Benetti & Branch & Wang & SN Name & \ion{C}{II} & SNID & Benetti & Branch & Wang \\
  & Type$^\textrm{a}$ & (Sub)Type$^\textrm{b}$ & Type$^\textrm{c}$ & Type$^\textrm{d}$ & Type$^\textrm{e}$ &  & Type$^\textrm{a}$ & (Sub)Type$^\textrm{b}$ & Type$^\textrm{c}$ & Type$^\textrm{d}$ & Type$^\textrm{e}$ \\
\hline
\endhead

\hline \hline
\multicolumn{12}{l}{Continued on Next Page\ldots} \\
\endfoot

\hline \hline
\endlastfoot

SN 1994D & A & Ia-norm & LVG & CN & N & SN 1998dm & A & Ia-norm & $\cdots$ & $\cdots$ & $\cdots$ \\
SN 2002cr & A & Ia-norm & $\cdots$ & $\cdots$ & $\cdots$ & SN 2002hw & A & Ia-norm & $\cdots$ & $\cdots$ & $\cdots$ \\
SN 2003kf & A & Ia-norm & $\cdots$ & $\cdots$ & $\cdots$ & SN 2004ey & A & Ia-norm & $\cdots$ & $\cdots$ & $\cdots$ \\
SN 2005cf & A & Ia-norm & HVG & CN & N & SN 2005el & A & Ia-norm & LVG & CN & N \\
SN 2005eu & A & Ia-norm & $\cdots$ & $\cdots$ & $\cdots$ & SN 2005iq & A & Ia-norm & $\cdots$ & $\cdots$ & $\cdots$ \\
SN 2005ki$^\textrm{f}$ & A & Ia-norm & LVG & BL & N & SN 2007F & A & Ia-norm & $\cdots$ & SS & N \\
SN 2007af$^\textrm{g}$ & A & Ia-norm & HVG & BL & N & SN 2007bm & A & Ia-norm & $\cdots$ & $\cdots$ & $\cdots$ \\
SN 2007cq & A & Ia & $\cdots$ & $\cdots$ & $\cdots$ & SN 2008s1$^\textrm{h}$ & A & Ia-norm & $\cdots$ & BL & N \\
\hline
SN 1995E & F & Ia-norm & $\cdots$ & CN & N & SN 1998dk & F & Ia-norm & $\cdots$ & $\cdots$ & HV \\
SN 1999dk & F & Ia-norm & $\cdots$ & $\cdots$ & $\cdots$ & SN 2000dn & F & Ia-norm & LVG & BL & N \\
SN 2001az & F & Ia-norm & $\cdots$ & $\cdots$ & N & SN 2001cp & F & Ia-norm & $\cdots$ & $\cdots$ & N \\
SN 2001fe & F & Ia-norm & $\cdots$ & SS & N & SN 2002ck & F & Ia-norm & $\cdots$ & CN & N \\
SN 2002er & F & Ia-norm & $\cdots$ & $\cdots$ & HV & SN 2003U & F & Ia-norm & $\cdots$ & BL & N \\
SN 2004fz & F & Ia-norm & $\cdots$ & $\cdots$ & $\cdots$ & SN 2005ao & F & Ia & $\cdots$ & SS & $\cdots$ \\
SN 2005na & F & Ia-norm & HVG & $\cdots$ & N & SN 2006ax & F & Ia-norm & $\cdots$ & $\cdots$ & $\cdots$ \\
SN 2006bt & F & Ia-norm & HVG & CL & N & SN 2006cs & F & Ia-91bg & $\cdots$ & CL & $\cdots$ \\
SN 2007bc & F & Ia-norm & $\cdots$ & CL & N & SN 2007on & F & Ia-norm & $\cdots$ & CL & N \\
SN 2008Z & F & Ia-99aa & $\cdots$ & $\cdots$ & $\cdots$ & SN 2008hs & F & Ia-norm & $\cdots$ & $\cdots$ & $\cdots$ \\
\hline
SN 1989M & N & Ia-norm & HVG & BL & HV & SN 1991bg & N & Ia-91bg & FAINT & $\cdots$ & $\cdots$ \\
SN 1994S & N & Ia-norm & $\cdots$ & SS & N & SN 1997Y & N & Ia-norm & $\cdots$ & BL & N \\
SN 1997br$^\textrm{i}$ & N & Ia-91T & $\cdots$ & $\cdots$ & $\cdots$ & SN 1997do & N & Ia-norm & $\cdots$ & $\cdots$ & $\cdots$ \\
SN 1998ef & N & Ia-norm & $\cdots$ & $\cdots$ & $\cdots$ & SN 1998es & N & Ia-99aa & $\cdots$ & SS & $\cdots$ \\
SN 1999aa & N & Ia-99aa & $\cdots$ & SS & $\cdots$ & SN 1999ac & N & Ia-norm & HVG & CN & N \\
SN 1999da & N & Ia-91bg & FAINT & CL & $\cdots$ & SN 1999dq & N & Ia-99aa & HVG & SS & $\cdots$ \\
SN 1999gd & N & Ia-norm & $\cdots$ & BL & N & SN 2000cp & N & Ia-norm & $\cdots$ & $\cdots$ & N \\
SN 2000dg & N & Ia-norm & $\cdots$ & SS & N & SN 2000dk & N & Ia-norm & FAINT & CL & N \\
SN 2000dm & N & Ia-norm & HVG & BL & N & SN 2001bf & N & Ia-norm & $\cdots$ & $\cdots$ & N \\
SN 2001br & N & Ia-norm & $\cdots$ & BL & HV & SN 2001da & N & Ia-norm & HVG & BL & N \\
SN 2001eh & N & Ia-99aa & $\cdots$ & SS & $\cdots$ & SN 2001ep & N & Ia-norm & HVG & CL & N \\
SN 2001ex & N & Ia-91bg & $\cdots$ & $\cdots$ & $\cdots$ & SN 2002aw & N & Ia-norm & $\cdots$ & $\cdots$ & N \\
SN 2002bf & N & Ia-norm & $\cdots$ & BL & HV & SN 2002bo & N & Ia-norm & HVG & $\cdots$ & HV \\
SN 2002cf & N & Ia-91bg & $\cdots$ & CL & $\cdots$ & SN 2002cs & N & Ia-norm & $\cdots$ & $\cdots$ & $\cdots$ \\
SN 2002cu & N & Ia-norm & $\cdots$ & $\cdots$ & $\cdots$ & SN 2002dj & N & Ia-norm & $\cdots$ & $\cdots$ & $\cdots$ \\
SN 2002dk & N & Ia-91bg & $\cdots$ & CL & $\cdots$ & SN 2002eb & N & Ia-norm & $\cdots$ & $\cdots$ & N \\
SN 2002eu & N & Ia-norm & HVG? & CL & N & SN 2002fb & N & Ia-91bg & $\cdots$ & CL & $\cdots$ \\
SN 2002ha & N & Ia-norm & LVG & BL & N & SN 2002he & N & Ia-norm & HVG & BL & HV \\
SN 2002hu & N & Ia-99aa & $\cdots$ & $\cdots$ & $\cdots$ & SN 2003W & N & Ia-norm & $\cdots$ & $\cdots$ & $\cdots$ \\
SN 2003cq & N & Ia-norm & $\cdots$ & $\cdots$ & HV & SN 2003gt & N & Ia-norm & $\cdots$ & $\cdots$ & $\cdots$ \\
SN 2003he & N & Ia-norm & LVG & BL & N & SN 2003iv & N & Ia-norm & HVG? & CL & N \\
SN 2004as & N & Ia-norm & $\cdots$ & BL & HV & SN 2004br & N & Ia-norm & $\cdots$ & $\cdots$ & $\cdots$ \\
SN 2004bv & N & Ia-91T & $\cdots$ & $\cdots$ & $\cdots$ & SN 2004bw & N & Ia-norm & $\cdots$ & $\cdots$ & N \\
SN 2004dt & N & Ia-norm & HVG & BL & HV & SN 2004ef & N & Ia-norm & $\cdots$ & $\cdots$ & HV \\
SN 2004eo & N & Ia-norm & $\cdots$ & $\cdots$ & $\cdots$ & SN 2004fu & N & Ia-norm & HVG? & BL & HV \\
SN 2004gs & N & Ia-norm & $\cdots$ & CL & N & SN 2005W & N & Ia-norm & $\cdots$ & BL & N \\
SN 2005ag$^\textrm{j}$ & N & Ia-norm & $\cdots$ & $\cdots$ & N & SN 2005bc & N & Ia-norm & LVG & CL & N \\
SN 2005de & N & Ia-norm & HVG & BL & N & SN 2005er & N & Ia-91bg & HVG? & CL & $\cdots$ \\
SN 2005eq & N & Ia-99aa & HVG & $\cdots$ & $\cdots$ & SN 2005lz & N & Ia-norm & $\cdots$ & BL & N \\
SN 2005ms & N & Ia-norm & HVG & $\cdots$ & N & SN 2006N & N & Ia-norm & HVG & BL & N \\
SN 2006S & N & Ia-norm & LVG & $\cdots$ & $\cdots$ & SN 2006X & N & Ia-norm & $\cdots$ & BL & HV \\
SN 2006bz & N & Ia-91bg & $\cdots$ & CL & $\cdots$ & SN 2006cj & N & Ia-norm & $\cdots$ & SS & N \\
SN 2006cp & N & Ia-norm & $\cdots$ & $\cdots$ & $\cdots$ & SN 2006cq & N & Ia-norm & $\cdots$ & $\cdots$ & N \\
SN 2006ef & N & Ia-norm & $\cdots$ & BL & HV & SN 2006ej & N & Ia-norm & HVG & BL & HV \\
SN 2006et & N & Ia-norm & LVG & SS & N & SN 2006gt & N & Ia-91bg & $\cdots$ & $\cdots$ & $\cdots$ \\
SN 2006ke & N & Ia-91bg & HVG? & $\cdots$ & $\cdots$ & SN 2006kf & N & Ia-norm & $\cdots$ & CL & N \\
SN 2006le & N & Ia-norm & $\cdots$ & $\cdots$ & $\cdots$ & SN 2006or & N & Ia-norm & $\cdots$ & BL & N \\
SN 2006sr & N & Ia-norm & HVG & BL & HV & SN 2007A & N & Ia-norm & LVG? & CN & N \\
SN 2007N & N & Ia & $\cdots$ & CL & $\cdots$ & SN 2007O & N & Ia-norm & $\cdots$ & SS & N \\
SN 2007bd & N & Ia-norm & $\cdots$ & $\cdots$ & N & SN 2007bz & N & Ia-norm & $\cdots$ & $\cdots$ & HV \\
SN 2007ca & N & Ia-norm & $\cdots$ & $\cdots$ & $\cdots$ & SN 2007ci & N & Ia-norm & $\cdots$ & CL & N \\
SN 2007co & N & Ia-norm & LVG & BL & N & SN 2007fb & N & Ia-norm & LVG? & $\cdots$ & N \\
SN 2007fr & N & Ia-norm & $\cdots$ & CL & N & SN 2007gi & N & Ia-norm & HVG? & $\cdots$ & HV \\
SN 2007gk & N & Ia-norm & HVG? & BL & HV & SN 2007hj & N & Ia-norm & FAINT & CL & HV \\
SN 2007le & N & Ia-norm & HVG & $\cdots$ & HV & SN 2007s1$^\textrm{k}$ & N & Ia-norm & $\cdots$ & BL & N \\
SN 2007qe & N & Ia-norm & $\cdots$ & $\cdots$ & HV & SN 2008ar & N & Ia-norm & $\cdots$ & CN & N \\
SN 2008ec & N & Ia-norm & LVG & CL & N & SN 2008ei & N & Ia-norm & HVG* & BL & HV \\
SN 2008s5$^\textrm{l}$ & N & Ia & LVG* & $\cdots$ & $\cdots$ &  & & & & & \\
\hline
SN 2000fa & ? & Ia-norm & $\cdots$ & $\cdots$ & N & SN 2002cd & ? & Ia-norm & LVG & $\cdots$ & HV \\
SN 2002de & ? & Ia-norm & HVG & CL & HV & SN 2003Y & ? & Ia-91bg & $\cdots$ & $\cdots$ & $\cdots$ \\
SN 2003gn & ? & Ia-norm & $\cdots$ & $\cdots$ & $\cdots$ & SN 2005dv & ? & Ia-norm & $\cdots$ & BL & HV \\
SN 2006cm & ? & Ia-norm & $\cdots$ & $\cdots$ & N & SN 2006cz & ? & Ia-99aa & $\cdots$ & $\cdots$ & $\cdots$ \\
SN 2006gr & ? & Ia-norm & $\cdots$ & $\cdots$ & $\cdots$ & SN 2006lf & ? & Ia-norm & $\cdots$ & $\cdots$ & $\cdots$ \\
SN 2007al & ? & Ia-91bg & $\cdots$ & $\cdots$ & $\cdots$ & SN 2008bt & ? & Ia-91bg & $\cdots$ & CL & $\cdots$ \\
SN 2008dx & ? & Ia-91bg & FAINT* & CL & $\cdots$ \\

\hline \hline
\multicolumn{12}{p{6.8in}}{$^\textrm{a}$Classification based on the presence or absence of \ion{C}{II} absorption in our spectra. `A' = clearly separated spectral absorption feature attributed to \ion{C}{II} is present; `F' = a flattening or depression in the red wing of the \ion{Si}{II} $\lambda$6355 feature is present that is likely due to \ion{C}{II}; `N' = no evidence for \ion{C}{II} is present; `?' = no definitive determination can be made about the presence or absence of \ion{C}{II} owing to noisy data.} \\
\multicolumn{12}{p{6.8in}}{$^\textrm{b}$Spectral classification using the SuperNova IDentification code \citep[SNID;][]{Blondin07} taken from Section~5 of BSNIP~I.} \\
\multicolumn{12}{p{6.8in}}{$^\textrm{c}$Classification based on the velocity gradient of the \ion{Si}{II} $\lambda$6355 line \citep{Benetti05}. `HVG' = high velocity gradient; `LVG' = low velocity gradient; `FAINT' = faint/underluminous. Classifications marked with a `?' are uncertain since light-curve shape information is unavailable. Classifications marked with a `*' use the MLCS2k2 $\Delta$ parameter \citep{Jha07} as a proxy for $\Delta m_{15}$. Taken from BSNIP~II.} \\
\multicolumn{12}{p{6.8in}}{$^\textrm{d}$Classification based on the pseudo-equivalent widths of the \ion{Si}{II} $\lambda$6355 and \ion{Si}{II} $\lambda$5972 lines \citep{Branch09}. `CN' = core normal; `BL' = broad line; `CL' = cool; `SS' = shallow silicon. Taken from BSNIP~II.} \\
\multicolumn{12}{p{6.8in}}{$^\textrm{e}$Classification based on the velocity of the \ion{Si}{II} $\lambda$6355 line \citep{Wang09}. `HV' = high velocity; `N' = normal. Taken from BSNIP~II.} \\
\multicolumn{12}{l}{$^\textrm{f}$\ion{C}{II} classification was changed from `N' based on the data presented by \citet{Thomas11}.} \\
\multicolumn{12}{l}{$^\textrm{g}$\ion{C}{II} classification was changed from `N' based on the data presented by \citet{Parrent11}.} \\
\multicolumn{12}{l}{$^\textrm{h}$Also known as SNF20080514-002.} \\
\multicolumn{12}{l}{$^\textrm{i}$\ion{C}{II} classification was changed from `?' based on the data presented by \citet{Parrent11}.} \\
\multicolumn{12}{l}{$^\textrm{j}$\ion{C}{II} classification was changed from `?' based on the data presented by \citet{Folatelli11}.} \\
\multicolumn{12}{l}{$^\textrm{k}$Also known as SNF20071021-000.} \\
\multicolumn{12}{l}{$^\textrm{l}$Also known as SNF20080909-030.} \\
\end{longtable}
\end{center}
\twocolumn
\setlength{\tabcolsep}{6pt}

\section{Carbon Detection and Measurement}\label{s:procedure}

As mentioned above, \ion{C}{II} is the dominant species of C for typical
SN~Ia temperatures before and near maximum brightness
\citep[\about10,000~K;
e.g.,][]{Hatano99}. This species has four major absorption lines in
the optical regime: $\lambda$4267, $\lambda$4745, $\lambda$6580,
and $\lambda$7234 \citep[e.g.,][]{Mazzali01,Branch03}. The bluest two
lines are usually overwhelmed in SN~Ia spectra by broad, 
blended absorption from iron-group elements (IGEs), so we do not
attempt to search for either of them in our
data. The $\lambda$7234 line is more promising, but still not
perfect since it is relatively weak, usually quite broad, and often
falls close to the telluric absorption feature at 6900~\AA\
\citep{Folatelli11}. When this line is detected, however,
it is often only observed as an ``inflection'' in the spectral
continuum \citep{Thomas11}, though it becomes more obvious when the
$\lambda$6580 line is easily detected \citep{Parrent11}.

The $\lambda$6580 line is the most obvious \ion{C}{II} absorption
feature in the optical \citep[e.g.,][]{Hatano99} and represents the
best chance of making a definitive detection of unburned C in
pre-maximum optical spectra of SNe~Ia. However, even though it is the
deepest \ion{C}{II} absorption, it is still significantly weaker
than the nearby \ion{Si}{II} $\lambda$6355 line. Furthermore,
\ion{C}{II} $\lambda$6580 is usually blueshifted to \about6300~\AA,
which often intersects the red wing (or emission component of a
P-Cygni profile) of the \ion{Si}{II} $\lambda$6355 line. Therefore,
even though we concentrate mainly on searching for \ion{C}{II}
$\lambda$6580, unambiguously observing this feature is still a
difficult task.

\subsection{The Search for Carbon}

The first step in our search for C consists of visually inspecting
each of the 188 spectra in our sample. We concentrate on the region
5600--7800~\AA\ in order to cover the spectral range around
\ion{Si}{II} $\lambda$6355 and \ion{C}{II} $\lambda$6580 as well as
\ion{C}{II} $\lambda$7234 and the \ion{O}{I} triplet (centered near
7770~\AA). Spectra where there is an obvious, distinct absorption
feature likely associated with \ion{C}{II} $\lambda$6580 is given an
`A' (``Absorption'') classification. `A' spectra often also show
depressions or distinct absorption features associated with
\ion{C}{II} $\lambda$7234. Spectra that show no distinct absorption,
but the possibility of a depression or flattening of the red wing of
the \ion{Si}{II} $\lambda$6355 feature are classified as `F'
(``Flattened''). These data represent tentative C detections, and no
`F' spectra show obvious evidence for \ion{C}{II} $\lambda$7234 
absorption. We consider any spectrum with an `A' or `F' classification
to ``have C'' or be ``C positive.''

`N' (``No C'') classifications are given to spectra
where there is no evidence of C absorption features and the red half
of the \ion{Si}{II} $\lambda$6355 line appears to be unaffected by
any other species. Finally, spectra where no definite
classification can be made (often due to low S/N; i.e., the noise
fluctuations were as large as possible C absorption features) are
denoted as `?' (``Inconclusive''). This classification scheme is
similar to those used previously \citep{Parrent11,Folatelli11}.

To more quantitatively determine a spectrum's C classification, we use
the spectrum-synthesis code {\tt SYNOW} \citep{Synow}.  {\tt SYNOW} is
a parametrised resonance-scattering code which allows for the
adjustment of chemical composition, optical depths, temperatures, and
velocities in order to 
help identify spectral features seen in SNe. We fit all spectra
initially classified as `A' or `F' using {\tt SYNOW}, both with and
without \ion{C}{II}, to investigate whether the addition of
\ion{C}{II} significantly improves the match between the synthetic and
observed spectra. Again, this is similar to previous work 
\citep{Parrent11,Folatelli11}, though \citet{Thomas11} use a different
spectral synthesis code, {\tt SYNAPPS} \citep*{Thomas11:synapps}.

\begin{figure}
\centering$
\begin{array}{c}
\includegraphics[width=3.4in]{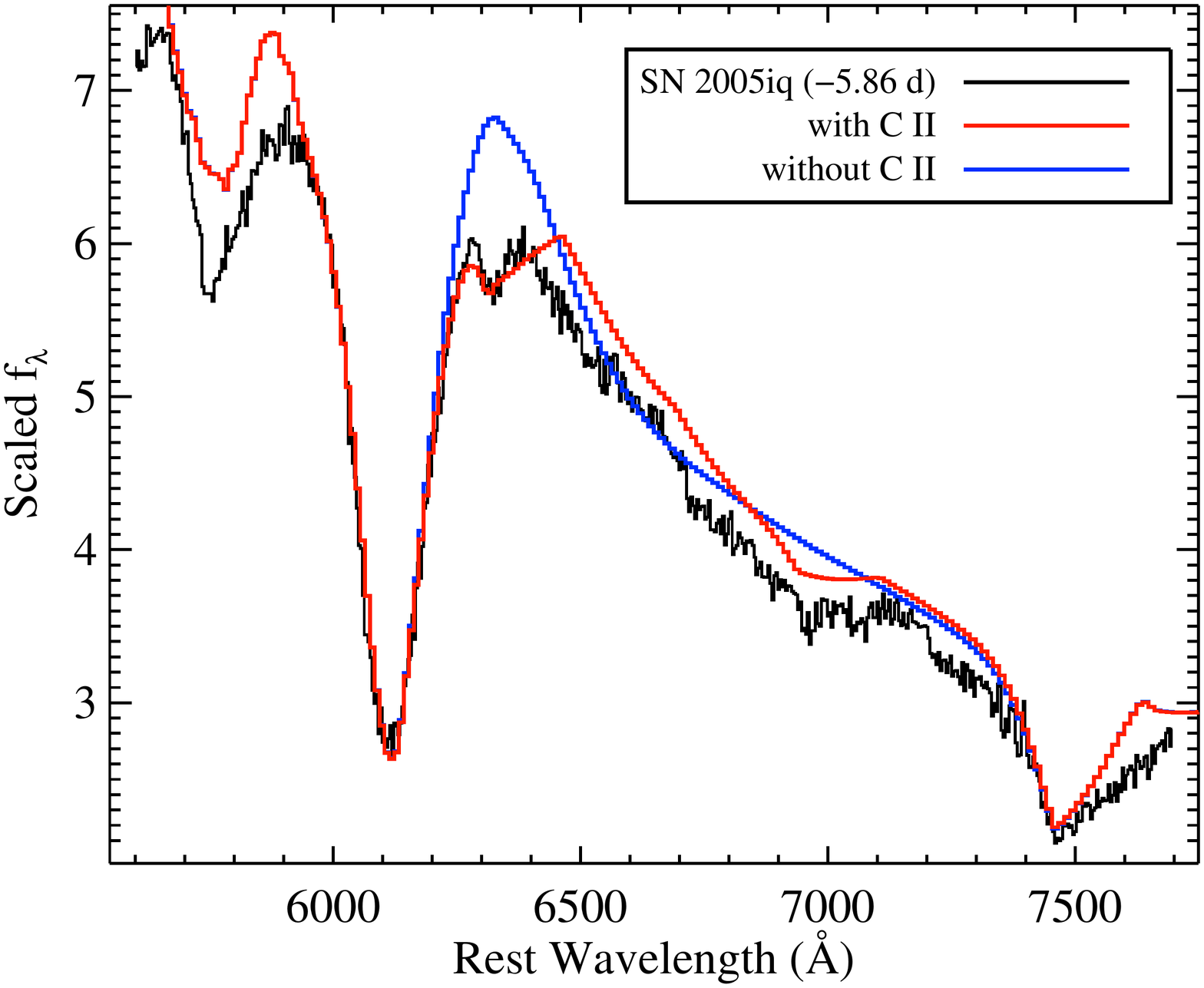} \\
\includegraphics[width=3.4in]{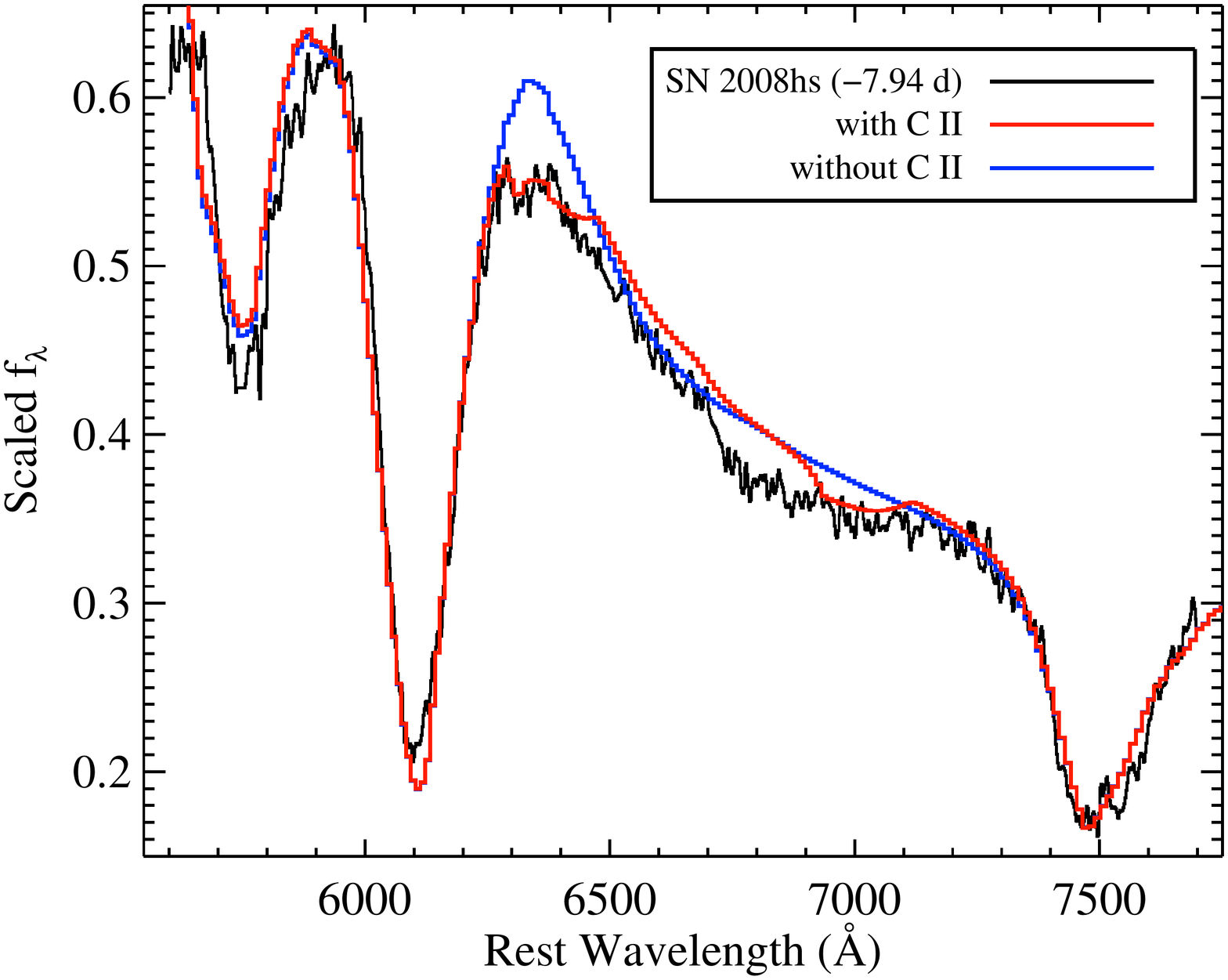} \\
\includegraphics[width=3.4in]{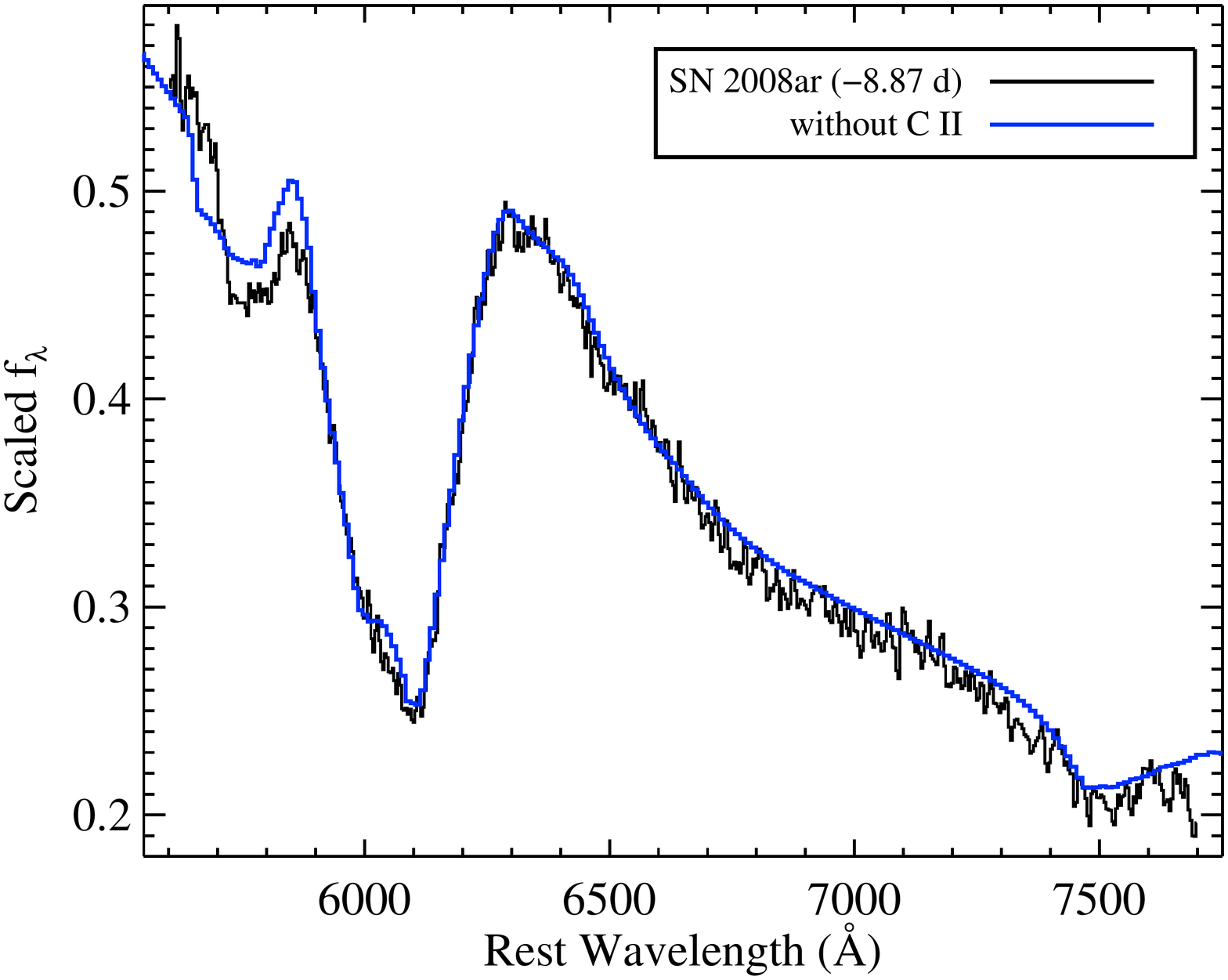} 
\end{array}$
\caption{Examples of observed and synthetic spectra (created using
  {\tt SYNOW}). ({\it Top}) SN~2005iq at 5.9~d before maximum brightness 
  is an `A' spectrum, and the synthetic spectrum is shown with and without
  \ion{C}{II} included (note that the two synthetic spectra are nearly
  identical at wavelengths below \about6200~\AA). ({\it Middle})
  SN~2008hs at 7.9~d before maximum 
  is an `F' spectrum, and again the synthetic spectrum is shown with
  and without \ion{C}{II}. ({\it Bottom}) SN~2008ar at 8.9~d before
  maximum is an `N' spectrum, and thus no \ion{C}{II} is included in
  the synthetic spectrum. The data are all deredshifted
  and dereddened using the redshift and reddening values presented in
  Table~1 of BSNIP~I, assuming that the extinction follows the
  \citet{Cardelli89} extinction law modified by
  \citet{ODonnell94}. }\label{f:synow_fits}
\end{figure}

After using
{\tt SYNOW} to fit all spectra thought to have C, it was
found that the addition of \ion{C}{II} did not improve the fit to 33
spectra that were initially classified as `F' (and thus they were
reclassified as `N'). This likely implies that the initial visual
inspections were perhaps a bit too ``optimistic'' in detecting
possible C absorptions. However, all spectra initially classified as
`A' were confirmed to contain C in their spectra from the {\tt SYNOW}
fits. Examples of observed and synthetic spectra of an `A' spectrum,
an `F' spectrum, and an `N' spectrum can be found in the top, middle,
and bottom panels of Figure~\ref{f:synow_fits}, respectively. 
Furthermore, Table~\ref{t:spectra} lists each spectrum in our sample, 
along with its age and carbon classification, and
a summary of the number of spectra in each class is presented in
Table~\ref{t:counts}. As mentioned above, the oldest spectrum that
shows evidence for \ion{C}{II} absorption (i.e., an `F'
classification) is \about3.6~d after maximum brightness.  
The oldest spectrum with an `A'
classification was obtained about 4.4~d before maximum.

\onecolumn
\begin{center}
\begin{longtable}{lrc|lrc|lrc}
\caption{Summary of Investigated Spectra}\label{t:spectra} \\[-2ex]
\hline \hline
SN Name & \multicolumn{1}{c}{Phase$^\textrm{a}$} & \ion{C}{II} & SN Name & \multicolumn{1}{c}{Phase$^\textrm{a}$} & \ion{C}{II} & SN Name & \multicolumn{1}{c}{Phase$^\textrm{a}$} & \ion{C}{II} \\
       &      & Type$^\textrm{b}$ &       &      & Type$^\textrm{b}$ &       &      & Type$^\textrm{b}$ \\
\hline
\endfirsthead
\multicolumn{9}{c}{{\tablename} \thetable{} --- Continued} \\
\hline \hline
SN Name & \multicolumn{1}{c}{Phase$^\textrm{a}$} & \ion{C}{II} & SN Name & \multicolumn{1}{c}{Phase$^\textrm{a}$} & \ion{C}{II} & SN Name & \multicolumn{1}{c}{Phase$^\textrm{a}$} & \ion{C}{II} \\
       &      & Type$^\textrm{b}$ &       &      & Type$^\textrm{b}$ &       &      & Type$^\textrm{b}$ \\
\hline
\endhead

\hline \hline
\multicolumn{9}{l}{Continued on Next Page\ldots} \\
\endfoot

\hline \hline
\endlastfoot

SN 1989M & $2.49$ & N & SN 1989M & $3.48$ & N & SN 1991bg & $0.14$ & N \\
SN 1991bg & $1.14$ & N & SN 1994D & $-12.31$ & A & SN 1994D & $-11.31$ & A \\
SN 1994D & $-9.32$ & A & SN 1994D & $-7.67$ & A & SN 1994D & $-5.32$ & A \\
SN 1994D & $-3.87$ & F & SN 1994D & $-3.33$ & N & SN 1994S & $1.11$ & N \\
SN 1995E & $-2.46$ & F & SN 1997Y & $1.27$ & N & SN 1997br & $-4.84$ & ? \\
SN 1997do & $-5.67$ & N & SN 1998dk & $-7.24$ & F & SN 1998dk & $-0.54$ & N \\
SN 1998dm & $-12.48$ & A & SN 1998dm & $-5.61$ & F & SN 1998ef & $-8.62$ & N \\
SN 1998es & $0.28$ & N & SN 1999aa & $0.24$ & N & SN 1999ac & $-3.70$ & N \\
SN 1999ac & $-0.89$ & N & SN 1999da & $-2.12$ & N & SN 1999dk & $-6.60$ & F \\
SN 1999dq & $-3.93$ & ? & SN 1999dq & $2.97$ & N & SN 1999gd & $-1.12$ & N \\
SN 2000cp & $2.92$ & N & SN 2000dg & $-5.09$ & N & SN 2000dk & $1.00$ & N \\
SN 2000dm & $-1.63$ & N & SN 2000dn & $-0.94$ & F & SN 2000fa & $-8.25$ & ? \\
SN 2001az & $-3.24$ & F & SN 2001bf & $1.22$ & N & SN 2001br & $3.47$ & ? \\
SN 2001br & $3.48$ & N & SN 2001cp & $1.39$ & F & SN 2001da & $-1.12$ & N \\
SN 2001eh & $3.26$ & N & SN 2001ep & $2.83$ & N & SN 2001ex & $-1.82$ & N \\
SN 2001fe & $-0.99$ & F & SN 2002aw & $2.10$ & N & SN 2002bf & $2.97$ & N \\
SN 2002bo & $-11.94$ & N & SN 2002bo & $-1.08$ & N & SN 2002cd & $1.10$ & ? \\
SN 2002cf & $-0.75$ & N & SN 2002ck & $3.64$ & F & SN 2002cr & $-6.78$ & A \\
SN 2002cs & $-7.76$ & N & SN 2002cu & $-5.28$ & N & SN 2002de & $-0.32$ & ? \\
SN 2002dj & $-7.98$ & N & SN 2002dk & $-1.23$ & N & SN 2002eb & $1.68$ & N \\
SN 2002er & $-4.58$ & F & SN 2002eu & $-0.06$ & N & SN 2002fb & $0.98$ & N \\
SN 2002ha & $-0.85$ & N & SN 2002he & $-1.03$ & N & SN 2002he & $0.29$ & N \\
SN 2002he & $3.22$ & N & SN 2002hu & $-5.81$ & N & SN 2002hw & $-6.27$ & A \\
SN 2003U & $-2.55$ & F & SN 2003W & $-5.06$ & N & SN 2003Y & $-1.74$ & ? \\
SN 2003cq & $-0.15$ & N & SN 2003gn & $-5.38$ & ? & SN 2003gt & $-5.07$ & N \\
SN 2003he & $2.71$ & N & SN 2003iv & $1.76$ & N & SN 2003kf & $-7.50$ & A \\
SN 2004as & $-4.36$ & N & SN 2004br & $3.50$ & N & SN 2004bv & $-7.06$ & N \\
SN 2004bw & $-10.03$ & N & SN 2004dt & $-6.46$ & ? & SN 2004dt & $1.38$ & N \\
SN 2004ef & $-5.52$ & N & SN 2004eo & $-5.57$ & N & SN 2004ey & $-7.58$ & A \\
SN 2004fu & $-2.65$ & N & SN 2004fu & $2.43$ & N & SN 2004fz & $-5.18$ & F \\
SN 2004gs & $0.44$ & N & SN 2005W & $0.59$ & N & SN 2005ag & $0.53$ & ? \\
SN 2005ao & $-1.29$ & F & SN 2005ao & $0.52$ & N & SN 2005bc & $1.55$ & N \\
SN 2005cf & $-10.94$ & A & SN 2005cf & $-2.11$ & F & SN 2005cf & $-1.19$ & N \\
SN 2005de & $-0.75$ & N & SN 2005dv & $-0.57$ & ? & SN 2005el & $-6.70$ & A \\
SN 2005el & $1.22$ & F & SN 2005er & $-0.26$ & ? & SN 2005er & $1.67$ & N \\
SN 2005eq & $-2.98$ & N & SN 2005eq & $0.66$ & ? & SN 2005eu & $-9.06$ & A \\
SN 2005eu & $-5.46$ & F & SN 2005iq & $-5.86$ & A & SN 2005ki & $1.62$ & N \\
SN 2005lz & $0.58$ & N & SN 2005ms & $-1.88$ & N & SN 2005na & $0.03$ & F \\
SN 2005na & $1.03$ & F & SN 2006N & $-1.89$ & N & SN 2006N & $-0.90$ & N \\
SN 2006S & $-3.93$ & ? & SN 2006S & $2.99$ & N & SN 2006X & $3.15$ & N \\
SN 2006ax & $-10.07$ & F & SN 2006bt & $-5.30$ & F & SN 2006bt & $-4.53$ & N \\
SN 2006bt & $2.27$ & N & SN 2006bz & $-2.44$ & N & SN 2006cj & $3.43$ & N \\
SN 2006cm & $-1.15$ & ? & SN 2006cp & $-5.30$ & N & SN 2006cq & $2.00$ & N \\
SN 2006cs & $2.28$ & F & SN 2006cz & $1.12$ & ? & SN 2006ef & $3.20$ & N \\
SN 2006gr & $-8.70$ & ? & SN 2006ej & $-3.70$ & N & SN 2006et & $3.29$ & N \\
SN 2006gt & $3.08$ & N & SN 2006ke & $2.36$ & N & SN 2006kf & $-8.96$ & ? \\
SN 2006kf & $-3.05$ & N & SN 2006lf & $-6.30$ & ? & SN 2006le & $-8.69$ & N \\
SN 2006or & $-2.79$ & N & SN 2006sr & $-2.34$ & N & SN 2006sr & $2.69$ & N \\
SN 2007A & $2.37$ & N & SN 2007F & $-9.35$ & A & SN 2007F & $3.23$ & N \\
SN 2007N & $0.44$ & N & SN 2007O & $-0.33$ & N & SN 2007af & $-1.25$ & N \\
SN 2007af & $2.84$ & N & SN 2007al & $3.39$ & ? & SN 2007bc & $0.61$ & F \\
SN 2007bd & $-5.79$ & N & SN 2007bm & $-7.79$ & A & SN 2007bz & $1.65$ & N \\
SN 2007ca & $-11.14$ & N & SN 2007ci & $-6.57$ & N & SN 2007ci & $-1.71$ & N \\
SN 2007co & $-4.09$ & N & SN 2007co & $0.85$ & N & SN 2007cq & $-5.82$ & A \\
SN 2007fb & $1.95$ & N & SN 2007fr & $-5.83$ & N & SN 2007fr & $-1.25$ & ? \\
SN 2007gi & $-7.31$ & N & SN 2007gi & $-0.35$ & N & SN 2007gk & $-1.72$ & N \\
SN 2007hj & $-1.23$ & N & SN 2007le & $-10.31$ & N & SN 2007le & $-9.40$ & N \\
SN 2007s1$^\textrm{c}$ & $-1.23$ & N & SN 2007on & $-3.01$ & F & SN 2007on & $-3.00$ & F \\
SN 2007qe & $-6.54$ & N & SN 2008Z & $-2.29$ & F & SN 2008ar & $-8.87$ & N \\
SN 2008ar & $2.83$ & N & SN 2008bt & $-1.08$ & ? & SN 2008s1$^\textrm{d}$ & $-6.36$ & A \\
SN 2008s1$^\textrm{d}$ & $-4.40$ & A & SN 2008s1$^\textrm{d}$ & $-3.42$ & F & SN 2008s1$^\textrm{d}$ & $0.49$ & F \\
SN 2008dx & $2.46$ & ? & SN 2008ec & $-0.24$ & N & SN 2008ei & $3.29$ & N \\
SN 2008s5$^\textrm{e}$ & $1.26$ & N & SN 2008hs & $-7.94$ & F \\
\hline \hline
\multicolumn{9}{p{5.5in}}{$^\textrm{a}$Phases of spectra are in rest-frame days relative to $B$-band maximum using the heliocentric redshift and photometry reference presented in Table~1 of BSNIP~I.} \\
\multicolumn{9}{p{5.5in}}{$^\textrm{b}$Classification based on the presence or absence of \ion{C}{II} absorption in our spectra. `A' = clearly separated spectral absorption feature attributed to \ion{C}{II} is present; `F' = a flattening or depression in the red wing of the \ion{Si}{II} $\lambda$6355 feature is present that is likely due to \ion{C}{II}; `N' = no evidence for \ion{C}{II} is present; `?' = no definitive determination can be made about the presence or absence of \ion{C}{II} owing to noisy data.} \\
\multicolumn{9}{l}{$^\textrm{c}$Also known as SNF20071021-000.} \\
\multicolumn{9}{l}{$^\textrm{d}$Also known as SNF20080514-002.} \\
\multicolumn{9}{l}{$^\textrm{e}$Also known as SNF20080909-030.} \\
\end{longtable}
\end{center}
\twocolumn
\setlength{\tabcolsep}{6pt}

\begin{table}
\begin{center}
\caption{Total Number of Objects and Spectra}\label{t:counts}
\begin{tabular}{lrr}
\hline\hline
C Class.$^\textrm{a}$ \hspace{.3in} & \multicolumn{1}{c}{\# Objects} & \# Spectra \\
\hline
A   &   16$^\textrm{b}$  &      19 \\
F     &  20\phantom{$^\textrm{b}$}     &  29 \\
N    &  95$^\textrm{c}$  &    117  \\
?      &  13\phantom{$^\textrm{b}$}     & 23 \\
\hline
Total  & 144\phantom{$^\textrm{b}$}      & 188 \\
\hline \hline
\multicolumn{3}{p{2.6in}}{$^\textrm{a}$Classification based on the
  presence or absence of \ion{C}{II} absorption in our spectra. `A' =
  clearly separated spectral absorption feature attributed to \ion{C}{II}
  is present; `F' = a flattening or depression in the red wing of the
  \ion{Si}{II} $\lambda$6355 feature is present that is likely due to
  \ion{C}{II}; `N' = no evidence for \ion{C}{II} is present; `?' = no
  definitive determination can be made about the presence or absence
  of \ion{C}{II} due to noisy data.} \\
\multicolumn{3}{p{2.6in}}{$^\textrm{b}$Two of these objects were classified
  as `N' using the BSNIP data, but reclassified as `A' based on data
  presented in previous work; see text.} \\
\multicolumn{3}{p{2.6in}}{$^\textrm{c}$Two of these objects were classified
  as `?' using the BSNIP data, but reclassified as `N' based on data
  presented in previous work; see text.} \\
\hline\hline
\end{tabular}
\end{center}
\end{table}

There are 34 SNe~Ia in the current sample for which we inspect
multiple spectra, and in many of these cases the C classifications of
different spectra of the same object disagree. However, this is
unsurprising since C features tend to weaken with time
\citep[e.g.,][]{Folatelli11}. In all cases using the BSNIP data, the
temporal evolution of the C classification is 
`A'$\rightarrow$`F'$\rightarrow$`N.' Therefore, we classify a SN by
the C classification of its earliest spectrum. Also, since any
spectrum classified as `?' does not indicate whether C is present, we
ignore any `?' spectra when determining the C classification of a
given SN.

When comparing the BSNIP data to previous studies, 33
SNe~Ia have been classified in earlier work
\citep{Parrent11,Thomas11,Folatelli11} and of these, our
classifications agree for 23 objects. For most of the objects where
the C classification differs, the BSNIP spectra are from earlier 
epochs or have higher S/N and show stronger evidence for C than
previously thought; for these objects we retain the C classification
determined from the BSNIP data.  However, there are two objects we
classify as `?' that were previously classified as `N' using higher
S/N data from earlier epochs, as well as two objects that we classify
as `N' that were previously classified as `A' using spectra from
earlier epochs. In these four cases we adopt the classification from
the literature in lieu of the one determined from our own data. These
reclassifications are reflected throughout this work. 
We note here that if we had access to more high-quality data or more
spectra at younger epochs, we might classify even more objects as C
positive. Therefore, the incidence rates of C in SNe~Ia that we
calculate herein should be considered a lower limit.

The final C classification for each SN~Ia in our sample (including
the above reclassifications) can be found in Table~\ref{t:objects}, 
and a summary of the number of objects in each class is presented in
Table~\ref{t:counts}. In the BSNIP data, \about11~per~cent of the 
SNe~Ia show definite C absorption features (`A'), while an additional 
\about14~per~cent show some evidence for C in their spectra (`F'),
for a total of \about25~per~cent `A' + `F.' This is consistent with 
previous studies that find \about20--33~per~cent of SNe~Ia show 
evidence for C \citep{Parrent11,Thomas11,Folatelli11}.

\subsection{Measuring the Carbon}\label{ss:measure}

\begin{figure*}
\centering
\centering$
\begin{array}{cc}
\includegraphics[width=3.35in]{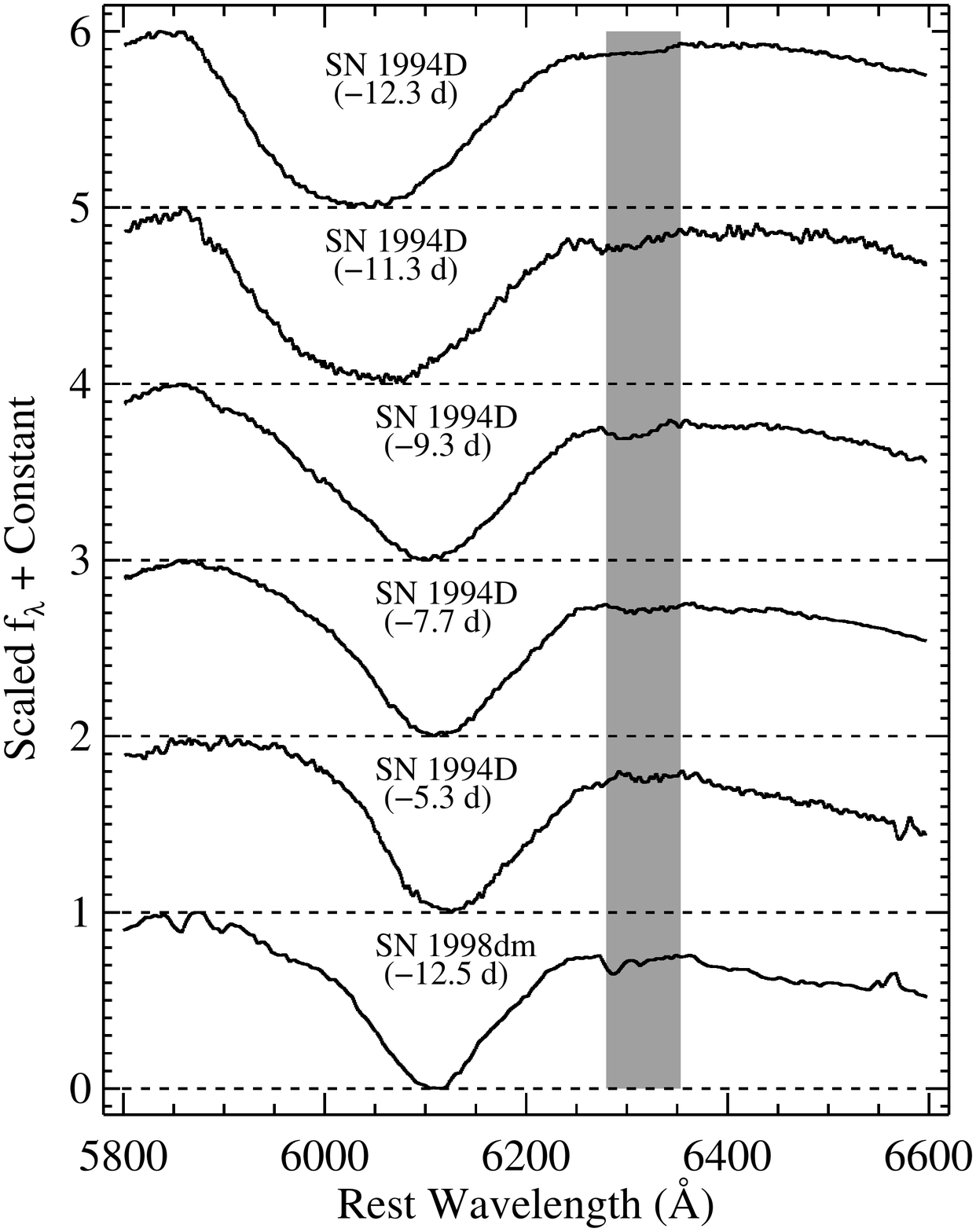} & 
\includegraphics[width=3.35in]{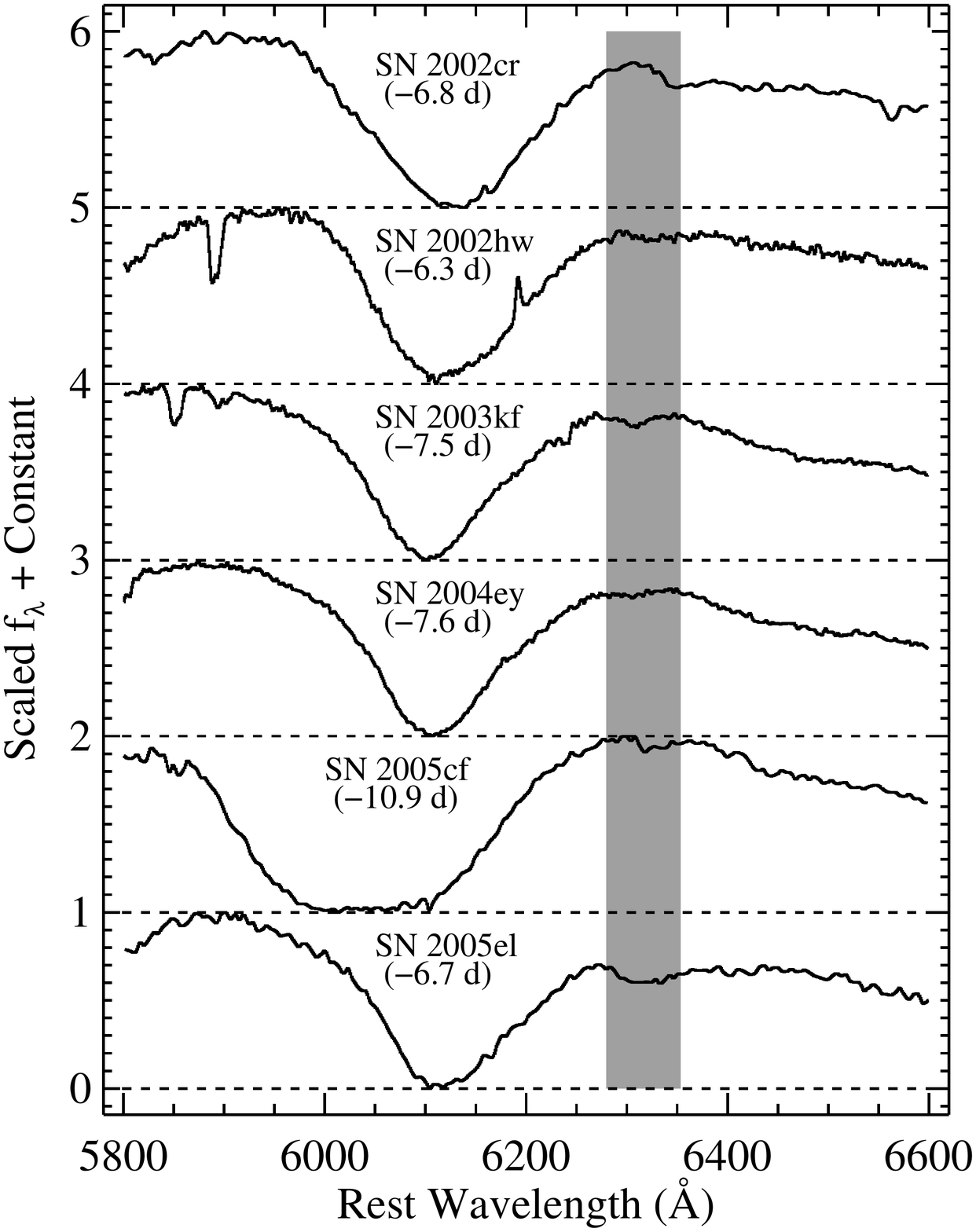} \\
\multicolumn{2}{c}{\includegraphics[width=3.35in]{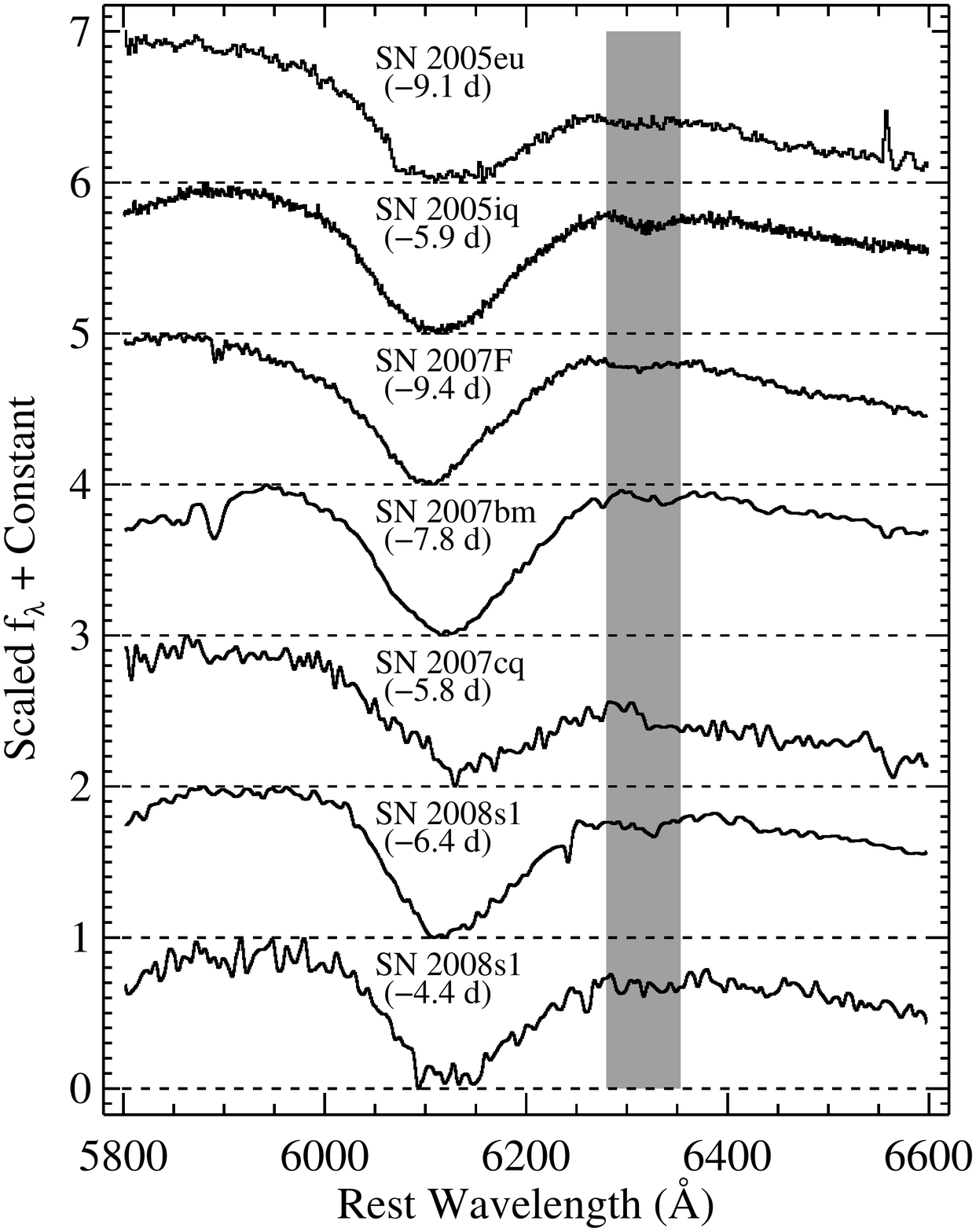}} \\
\end{array}$
\caption{All 19 of our `A' spectra showing the region near \ion{C}{II}
  $\lambda$6580 and \ion{Si}{II} $\lambda$6355. The grey regions cover
  wavelengths corresponding to \ion{C}{II} $\lambda$6580 with
  expansion velocities of 10,500--14,000~\kms\ (i.e., the range of
  velocities observed in this work with SNe~2002cr and 1994D having
  the smallest and largest velocities, respectively). Each spectrum is
  labeled with the SN name and rest-frame age. The data are all
  deredshifted and dereddened using the redshift and reddening values
  presented in Table~1 of BSNIP~I, assuming that the extinction
  follows the \citet{Cardelli89} extinction law modified by
  \citet{ODonnell94}.}\label{f:carbons}
\end{figure*}

Figure~\ref{f:carbons} shows the region
near \ion{C}{II} $\lambda$6580 
and \ion{Si}{II} $\lambda$6355 of all 19 spectra in our sample that are
classified as `A.' The wavelengths corresponding to \ion{C}{II}
$\lambda$6580 with expansion
velocities of 10,500--14,000~\kms\ (i.e., the range of velocities
observed in this work) are highlighted. 
For each of these spectra, we determine the pEW and
expansion velocity of the \ion{C}{II} $\lambda$6580 feature, and we
attempt to measure these parameters for the \ion{C}{II} $\lambda$7234
feature as well. The algorithm used to measure the \ion{C}{II}
$\lambda$6580 absorption is described in detail in BSNIP~II, but here
we give a brief summary of the procedure.

Each spectrum first has its host-galaxy recession velocity removed and is 
corrected for Galactic reddening (according to the values presented in 
Table~1 of BSNIP~I), and then is smoothed using a Savitzky-Golay smoothing
filter \citep{Savitzky64}. We attempt to define a pseudo-continuum for
each spectral feature. This is done by determining where the local
slope changes sign on either side of the feature's minimum. Quadratic
functions are fit to each of these endpoints, and the peaks of the
parabolas (assuming that they are both concave downward) are used as
the endpoints of the feature; they are then connected with a line to
define the pseudo-continuum. This defines the pEW
\citep[e.g.,][]{Garavini07}. Once a pseudo-continuum is calculated, a
cubic spline is fit to the smoothed data between the endpoints of the
spectral feature. The expansion velocity is calculated from the 
wavelength at which the spline fit reaches its minimum. Every fit is
visually inspected, and the fits to all of the 19 `A' spectra are found
to be acceptable.

We attempt to use the same procedure for the \ion{C}{II}
$\lambda$7234 absorption feature, but due to its relative shallowness
and the difficulty in automatically defining the endpoints, our
fitting routine fails on nearly all of the `A' spectra. Therefore, a
manual version of the algorithm is used (i.e., the endpoints are defined
by hand). Even so, the $\lambda$7234 absorption cannot be
accurately measured in 7 of 19 `A' spectra. Also, as a sanity check,
this manual version of the fitting procedure is used to measure the
$\lambda$6580 feature in all of the `A' spectra and the results are
compared to those of the more robust, automated fitting method
described above. All values of pEW and velocity from these two methods
are consistent within
the measured uncertainties, adding credibility to the values
measured for the $\lambda$7234 feature. Note that throughout the
analysis presented here we use only pEW and velocity measurements for
the $\lambda$6580 feature as determined by our automated fitting
routine. The results of these measurements, for both \ion{C}{II}
features inspected, can be found in Table~\ref{t:vels}.

\begin{table*}
\begin{center}
\caption{Measured Values for \ion{C}{II} Lines}\label{t:vels}
\begin{tabular}{lrrrrr}
\hline\hline
SN Name & \multicolumn{1}{c}{Phase$^\textrm{a}$} & \multicolumn{1}{c}{$\lambda$6580 pEW$^\textrm{b}$} & \multicolumn{1}{c}{$\lambda$6580 $v$$^\textrm{c}$} & \multicolumn{1}{c}{$\lambda$7234 pEW$^\textrm{b}$} & \multicolumn{1}{c}{$\lambda$7234 $v$$^\textrm{c}$} \\
\hline
SN 1994D & $-12.31$ & 12.22 (0.09) & 0.5 (0.1) & 11.23 (0.11) & 10.2 (0.5) \\
SN 1994D & $-11.31$ & 13.88 (0.10) & 2.6 (0.3) & 11.44 (0.11) & 10.5 (0.5) \\
SN 1994D & $-9.32$ & 13.03 (0.09) & 1.4 (0.2) & 10.11 (0.10) & 6.9 (0.3) \\
SN 1994D & $-7.67$ & 12.65 (0.09) & 0.9 (0.1) & 8.91 (0.09) & 6.5 (0.3) \\
SN 1994D & $-5.32$ & 12.27 (0.09) & 0.6 (0.1) & $\cdots$ & $\cdots$ \\
SN 1998dm & $-12.48$ & 13.49 (0.10) & 1.7 (0.2) & 11.27 (0.11) & 6.2 (0.3) \\
SN 2002cr & $-6.78$ & 10.72 (0.16) & 2.1 (0.3) & 9.60 (0.10) & 6.0 (0.3) \\
SN 2002hw & $-6.27$ & 12.14 (0.14) & 1.0 (0.2) & $\cdots$ & $\cdots$ \\
SN 2003kf & $-7.50$ & 12.72 (0.16) & 0.8 (0.1) & $\cdots$ & $\cdots$ \\
SN 2004ey & $-7.58$ & 12.94 (0.09) & 0.2 (0.1) & $\cdots$ & $\cdots$ \\
SN 2005cf & $-10.94$ & 11.88 (0.09) & 1.0 (0.2) & $\cdots$ & $\cdots$ \\
SN 2005el & $-6.70$ & 12.50 (0.09) & 3.1 (0.4) & 10.24 (0.10) & 7.2 (0.4) \\
SN 2005eu & $-9.06$ & 12.88 (0.37) & 0.8 (0.1) & $\cdots$ & $\cdots$ \\
SN 2005iq & $-5.86$ & 12.02 (0.17) & 2.7 (0.5) & 11.31 (0.11) & 4.1 (0.2) \\
SN 2007F & $-9.35$ & 12.73 (0.16) & 1.1 (0.2) & 11.01 (0.11) & 4.3 (0.2) \\
SN 2007bm & $-7.79$ & 11.25 (0.09) & 1.5 (0.2) & 9.47 (0.09) & 6.8 (0.3) \\
SN 2007cq & $-5.82$ & 11.82 (0.17) & 1.9 (0.3) & 9.47 (0.09) & 4.7 (0.2) \\
SN 2008s1$^\textrm{d}$ & $-6.36$ & 11.82 (0.16) & 1.9 (0.3) & 11.18 (0.11) & 4.0 (0.2) \\
SN 2008s1$^\textrm{d}$ & $-4.40$ & 11.00 (0.09) & 1.7 (0.4) & $\cdots$ & $\cdots$ \\
\hline \hline
\multicolumn{6}{l}{1$\sigma$ uncertainties for each measured value are given in parentheses.} \\
\multicolumn{6}{p{4in}}{$^\textrm{a}$Phases of spectra are in rest-frame days using the heliocentric redshift and photometry reference presented in Table~1 of BSNIP~I.} \\
\multicolumn{6}{l}{$^\textrm{b}$The pEW is in units of \AA.} \\
\multicolumn{6}{l}{$^\textrm{c}$The expansion velocity is in units of 1000 km~s$^{-1}$.} \\
\multicolumn{6}{l}{$^\textrm{d}$Also known as SNF20080514-002.} \\
\hline\hline
\end{tabular}
\end{center}
\end{table*}

\section{Analysis}\label{s:analysis}

\subsection{When is Carbon Detectable?}\label{ss:time}

As stated above, the oldest `A' spectrum in the BSNIP sample is from
4.4~d {\it before} maximum brightness and the oldest `F' spectrum was 
obtained 3.6~d {\it after} maximum. This matches well with previous
studies, which found that `A' spectra are found at ages less than 3~d before 
maximum \citep{Parrent11,Thomas11,Folatelli11}. Furthermore, all three
of the earlier investigations only use pre-maximum spectra (for
SNe~Ia that are not possible super-Chandrasekhar-mass or
SN~2002cx-like objects), and like the current study they find `F'
spectra near maximum brightness. 

\begin{figure}
\centering$
\begin{array}{c}
\includegraphics[width=3.4in]{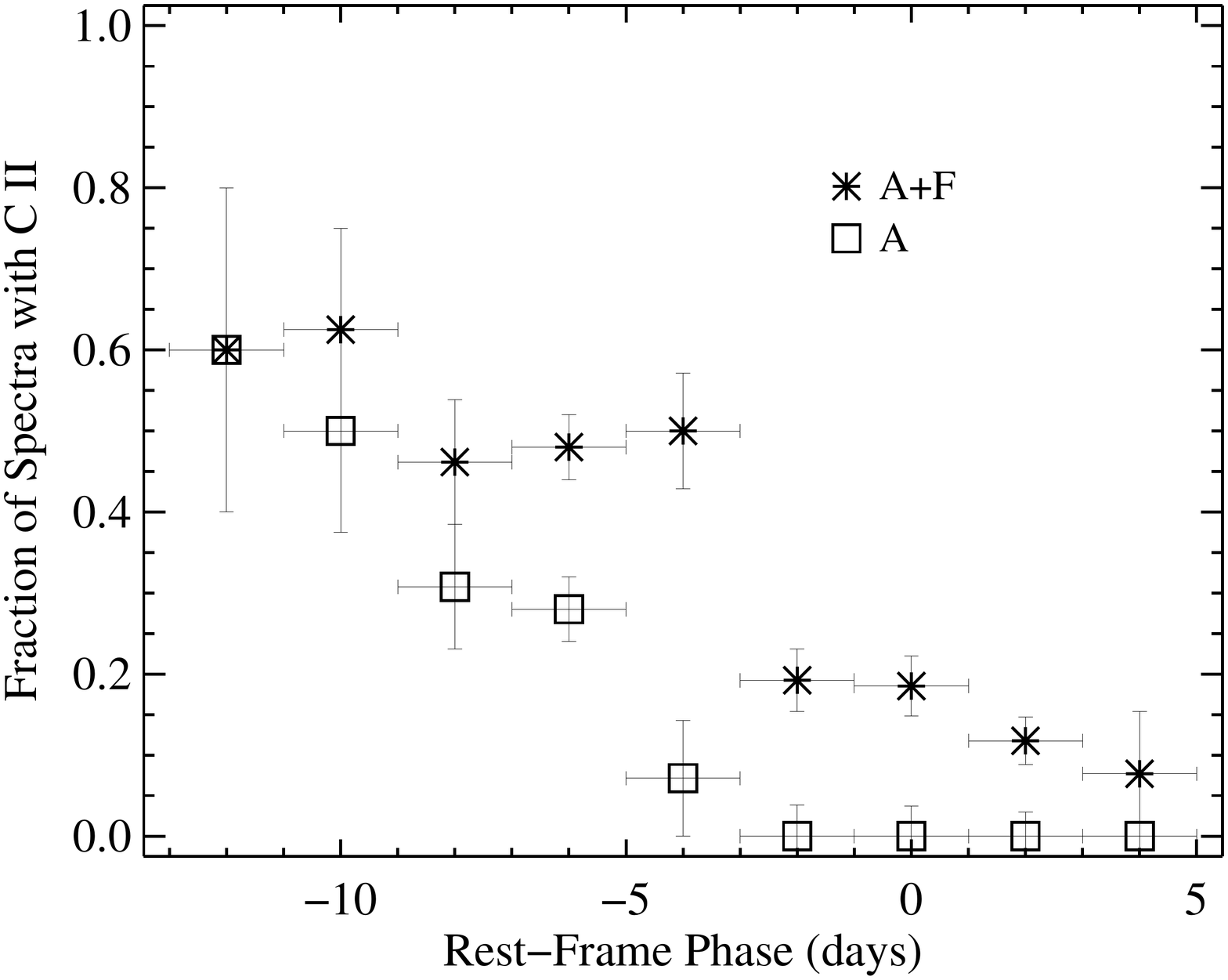} \\
\includegraphics[width=3.4in]{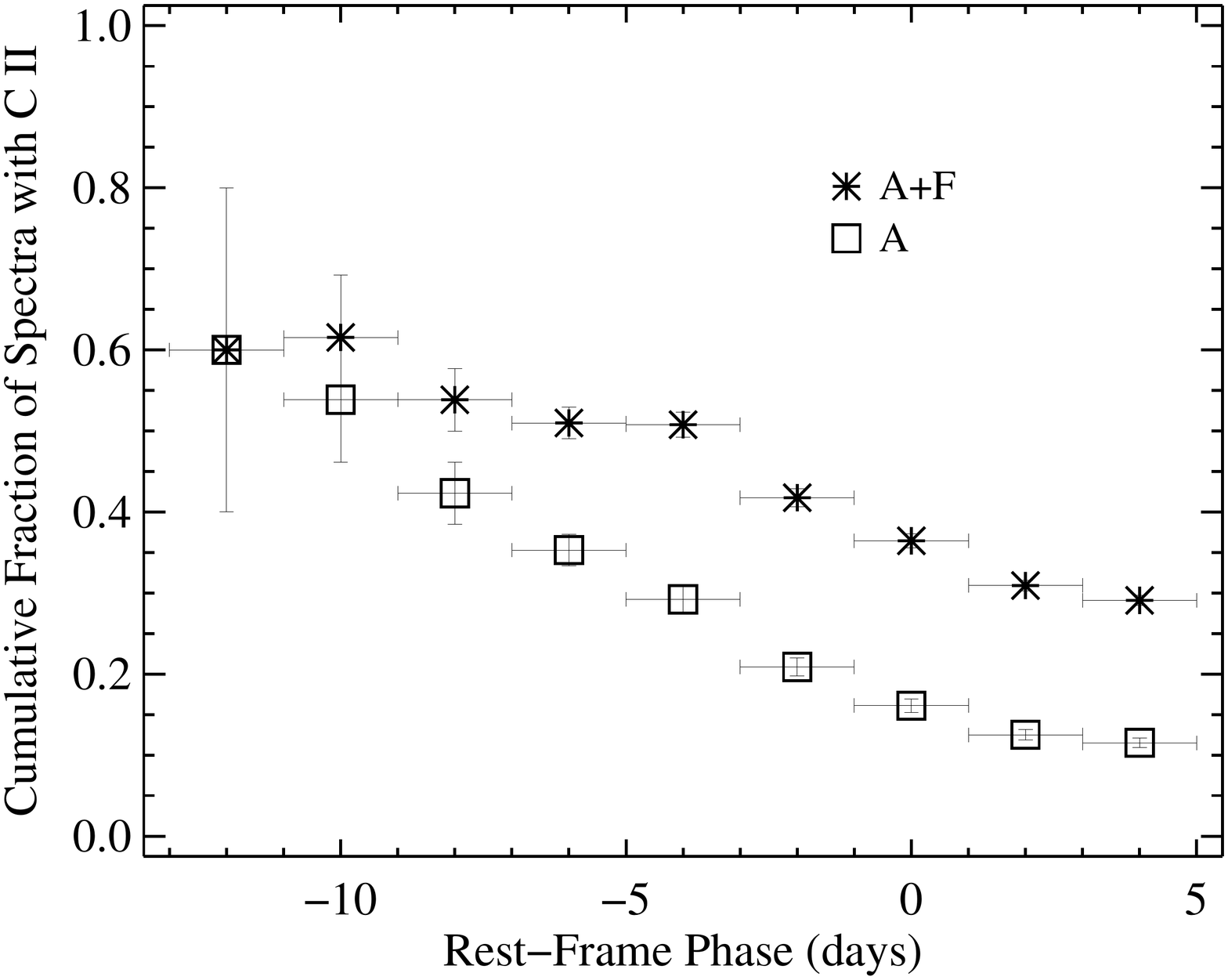} \\
\end{array}$
\caption{({\it Top}) The fraction of spectra with C versus time. The
  sum of `A' and `F' are asterisks and `A' alone are squares. The
  horizontal error bars represent the width of each bin (i.e., 2~d)
  and the vertical error bars represent the range of fractions if one
  SN with C is added to or subtracted from that bin. ({\it Bottom}) The 
  {\it cumulative} fraction of spectra with C versus time. This is the 
  fraction of spectra with C in that age bin or
  younger. Symbols and error bars have the same meanings as in the top 
  panel.}\label{f:fractions} 
\end{figure}

The top panel of Figure~\ref{f:fractions} shows the fraction of
spectra with C (just `A,' and the sum of `A' and `F')  as a function of
time. The horizontal error bars represent the width of each bin
(i.e., 2~d) and the vertical error bars represent the range of
fractions if one SN with C is added to or subtracted from that
bin. The fraction of `A' spectra and `A'+`F' spectra both start at
60~per~cent at 12~d before maximum brightness. The fraction of `A' spectra
decreases monotonically with time, while the fraction of `A'+`F'
spectra generally (but not always) decreases with time as well.

The small rise in the fraction of `A'+`F' spectra from $t = -12$~d to $t =
-10$~d appears to be insignificant. However, the rise of `A'+`F'
spectra for $-8\textrm{ d} \leq t \leq -4$~d seems inconsistent with
an actual decrease. It is unclear  what might cause this small spike
in the fraction of spectra with C at these epochs. This is consistent
with what was seen by \citet{Folatelli11} in their Figure~11, but not
exactly the same. Their fractions of `A' and `A'+`F' both
monotonically decrease with time, and they detect no `F' spectra older
than 2~d before maximum. Using the BSNIP data, 10--20~per~cent of the
spectra with $-2\textrm{~d} < t < 5$~d are classified as `F.'

The {\it cumulative} fraction of spectra with C (both `A' and `A'+`F')
is shown as a function of time in the bottom panel of
Figure~\ref{f:fractions}. Each point represents the fraction of
spectra with C from epochs in that bin or younger. The symbols and
error bars have the same meanings as in the top panel. Once again
there is a monotonic decrease in the fraction of `A' spectra with time,
in addition to mostly decreasing fractions of C-positive spectra (i.e.,
`A'+`F') with time. By $t \approx 5$~d, which corresponds to the age
bin that includes our oldest `F' spectrum, \about12~per~cent of the
spectra in the BSNIP sample show definitive C signatures (i.e., `A')
while \about29~per~cent of them show at least possible evidence for C
(i.e., `A'+`F').

It seems that if one wants to detect C in an optical spectrum of a
SN~Ia that follows the Phillips relation, a relatively high-quality 
spectrum must be obtained at an epoch younger than \about4~d past maximum
brightness. However, observations at the end of this epoch range will
yield only a possible C 
signature (i.e., an `F' spectrum) and will occur $<$10~per~cent of
the time. At $t \approx -4$~d, the chance of detecting C goes up
significantly. For the BSNIP data there is about a 50~per~cent chance
of obtaining an `A' or `F' spectrum at this epoch, though the
probability of obtaining an `A' spectrum is still only
\about8~per~cent. Finally, it appears that obtaining spectra at $t \la
-5$~d yields a relatively good chance of showing some sign of C (just
over 50~per~cent) and a better than one-third chance of yielding an
`A' spectrum.

\subsection{Carbon and Various Classification Schemes}\label{ss:classification}

In Table~\ref{t:objects}, the ``Carbon Classification'' for each
object studied here is listed, along with its spectral
classification based on various other classification methods. The
``SNID type'' of each SN is taken from BSNIP~I. The
SuperNova IDentification code \citep[SNID;][]{Blondin07}, as
implemented in BSNIP~I, was used to determine the spectroscopic
subtype of each SN in the BSNIP sample. SNID compares an input
spectrum to a library of spectral templates in order to determine the
most likely spectroscopic subtype. Spectroscopically normal objects
are objects classified as ``Ia-norm'' by SNID.

The spectroscopically peculiar SNID subtypes used here include the
often underluminous SN~1991bg-like objects \citep[``Ia-91bg,''
e.g.,][]{Filippenko92:91bg,Leibundgut93}, and the often overluminous
SN~1991T-like objects \citep[``Ia-91T,''
e.g.,][]{Filippenko92:91T,Phillips92} and SN~1999aa-like objects
\citep[``Ia-99aa,''][]{Li01:pec,Strolger02,Garavini04}. See BSNIP~I
for more information regarding our implementation of SNID and the
various spectroscopic subtype classifications. If an object has a SNID
type of simply ``Ia,'' it means that no definitive subtype could be
determined. 

According to Table~\ref{t:objects}, all `A' objects are Ia-norm,
with the exception of one ``Ia.'' All `F' objects are also Ia-norm,
except one each of Ia-99aa, Ia-91bg, and Ia. On the other hand, all
SNID types are well represented (relative to their overall incidence
rate) in the `N' objects. However, due to the relative rarity of the
spectroscopically peculiar subtypes, one would expect only 1--2 of
each of the non-Ia-norm subtypes in a sample of 19 objects
\citep{Ganeshalingam10:phot_paper,Li11a}. Thus, it seems that the
SNe~Ia showing evidence for C in their spectra are spectroscopically
normal objects when examining their entire optical spectrum (as SNID
does). Note, though, that there {\it could} exist very rare
cases of spectroscopically peculiar SNe~Ia that follow the Phillips
relation and that show C features.

The fourth column of Table~\ref{t:objects} presents the ``Benetti
type'' of each object, which is based on the velocity gradient of the
\ion{Si}{II} $\lambda$6355 feature \citep{Benetti05}. The high
velocity gradient (HVG) group has the largest velocity gradients while
the low velocity gradient (LVG) group has the smallest velocity
gradients. The third subclass (FAINT) has the lowest expansion
velocities, yet moderately large velocity gradients, and consists of
subluminous SNe~Ia with the narrowest light curves. As mentioned in 
BSNIP~II, from where these classifications are taken, the BSNIP data
are not well suited to velocity-gradient measurements since the average
number of spectra per object is \about2 (see BSNIP~I). However, we are
still able to calculate the velocity gradient for a subset of our
data.

Of the SNe~Ia with C and a known Benetti type, four are LVG (three of which 
are `A') and four are HVG (two of which are `A'). We also find that the actual
values of the velocity gradient itself are similar for objects with
and without C. \citet{Parrent11} find that
LVG objects have a greater chance of showing C as compared to HVG
objects and that no HVG SNe show a definitive C signature. They point
out that this could be an observational bias since HVG objects tend to
have higher \ion{Si}{II} velocities
\citep[e.g.,][]{Hachinger06,Wang09}, increasing the amount of
blending between \ion{Si}{II} $\lambda$6355 and \ion{C}{II}
$\lambda$6580 and thus making it more difficult to detect C. In BSNIP~II
we show that the one-to-one association between HVG and high expansion
velocities is not as clear as has been assumed previously, which could
explain how we are able to detect C in some HVG objects. However, this
possible connection between velocity gradient and incidence of C
should be explored further in the future with datasets that are more
suited than BSNIP to velocity-gradient calculations.

The ``Branch type'' referred to in Table~\ref{t:objects} uses pEWs of 
\ion{Si}{II} $\lambda$6355 and \ion{Si}{II} $\lambda$5972 measured
near maximum light to classify SNe~Ia \citep{Branch06}. The four groups
they define based on these two pEW values are core normal (CN), broad
line (BL), cool (CL), and shallow silicon (SS). However, they point
out that SNe seem to have a continuous distribution of pEW values;
hence, how the exact boundaries are defined is not critical. The
Branch-type classifications used here can be found in BSNIP~II. The
majority (63~per~cent) of CN objects show evidence for C, while only
16~per~cent of BL objects have C in their spectra. This is consistent
with the idea mentioned above that it is harder to distinguish C
absorption in SNe~Ia having high expansion velocities such as
BL objects \citep{Parrent11,Folatelli11}. 

Furthermore, only 25~per~cent and 18~per~cent of SS and CL objects
show evidence for C, respectively, which has been noticed 
in earlier work \citep{Parrent11,Folatelli11}. This has been
interpreted as evidence that the presence or absence of C depends on the
effective temperature \citep{Parrent11}, since the effective
temperature is directly related to the relative strengths of the
\ion{Si}{II} $\lambda$6355 and \ion{Si}{II} $\lambda$5972 features 
\citep{Nugent95}.

The prevalence of CN objects with C, as compared to other Branch
types, is unsurprising given what was found above when discussing SNID
types. In BSNIP~II it was shown that SNID types are often equivalent
to the more extreme objects in each of the non-CN Branch
types. Therefore, since nearly all SNe~Ia with C are Ia-norm, it
stands to reason that most of them should also be CN.

The last column of Table~\ref{t:objects} lists the ``Wang type'' of
each object, taken from BSNIP~II, which is determined from the
\ion{Si}{II} $\lambda$6355 velocity near maximum brightness 
\citep{Wang09}. SNe~Ia which are classified as Ia-norm by SNID and
have high velocities near maximum ($\ga$11,800~\kms) are considered
to be high-velocity (HV) objects. Ia-norm with velocities less than
this cutoff are classified as normal (N).

No HV objects are classified as `A' and only two are classified as `F,'
while 21 HV objects show no evidence of C. Nearly one-third of 
normal-velocity objects, however, have C. This is consistent with the
relative lack of BL objects that show C, since both BL and HV SNe have large
expansion velocities and this makes C detection difficult
\citep{Parrent11,Folatelli11}.

\subsection{Similarities (and Differences) Between Objects With and
  Without Carbon}\label{ss:comparison}

\begin{figure}
\centering$
\begin{array}{c}
\includegraphics[width=3.4in]{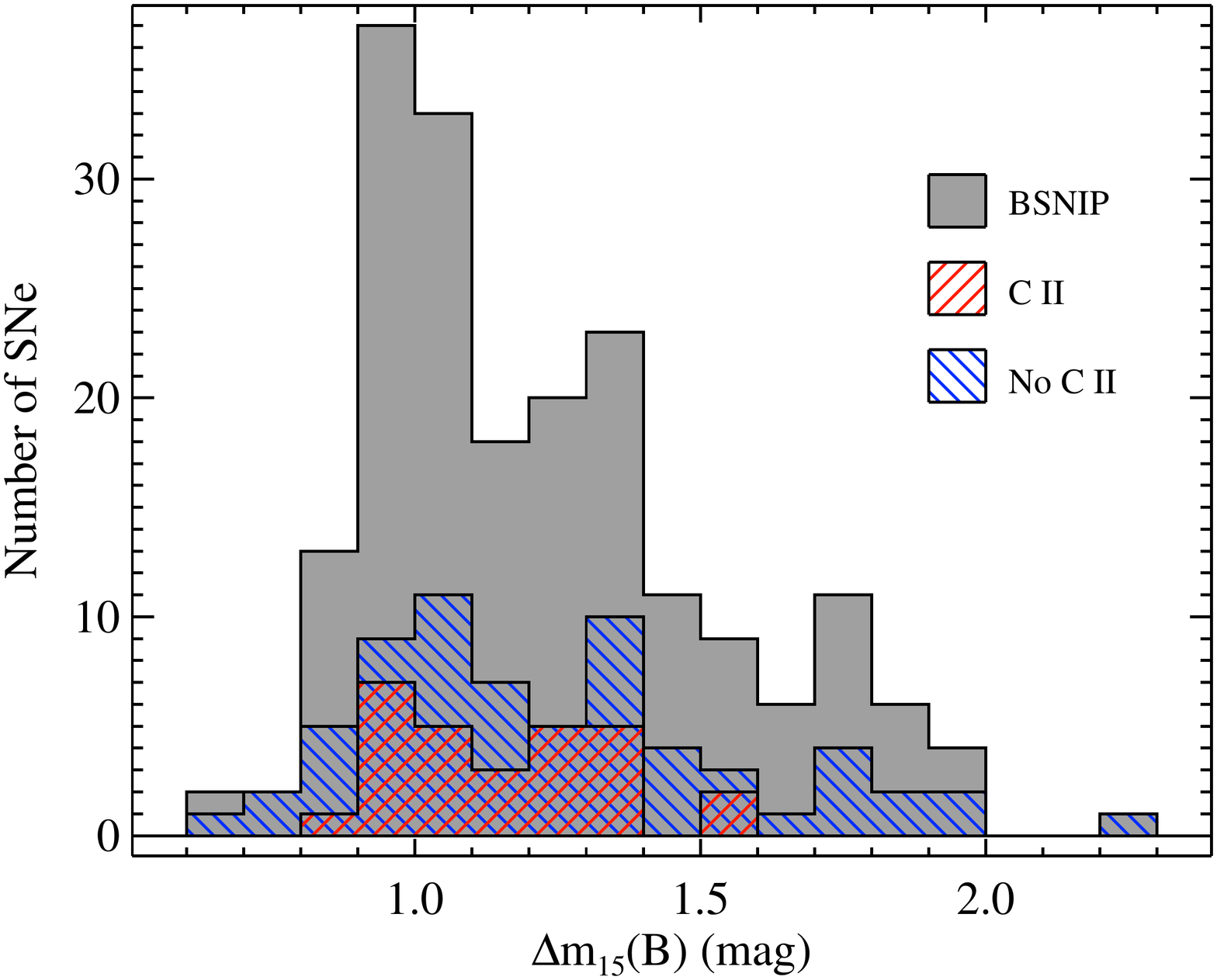} \\
\includegraphics[width=3.4in]{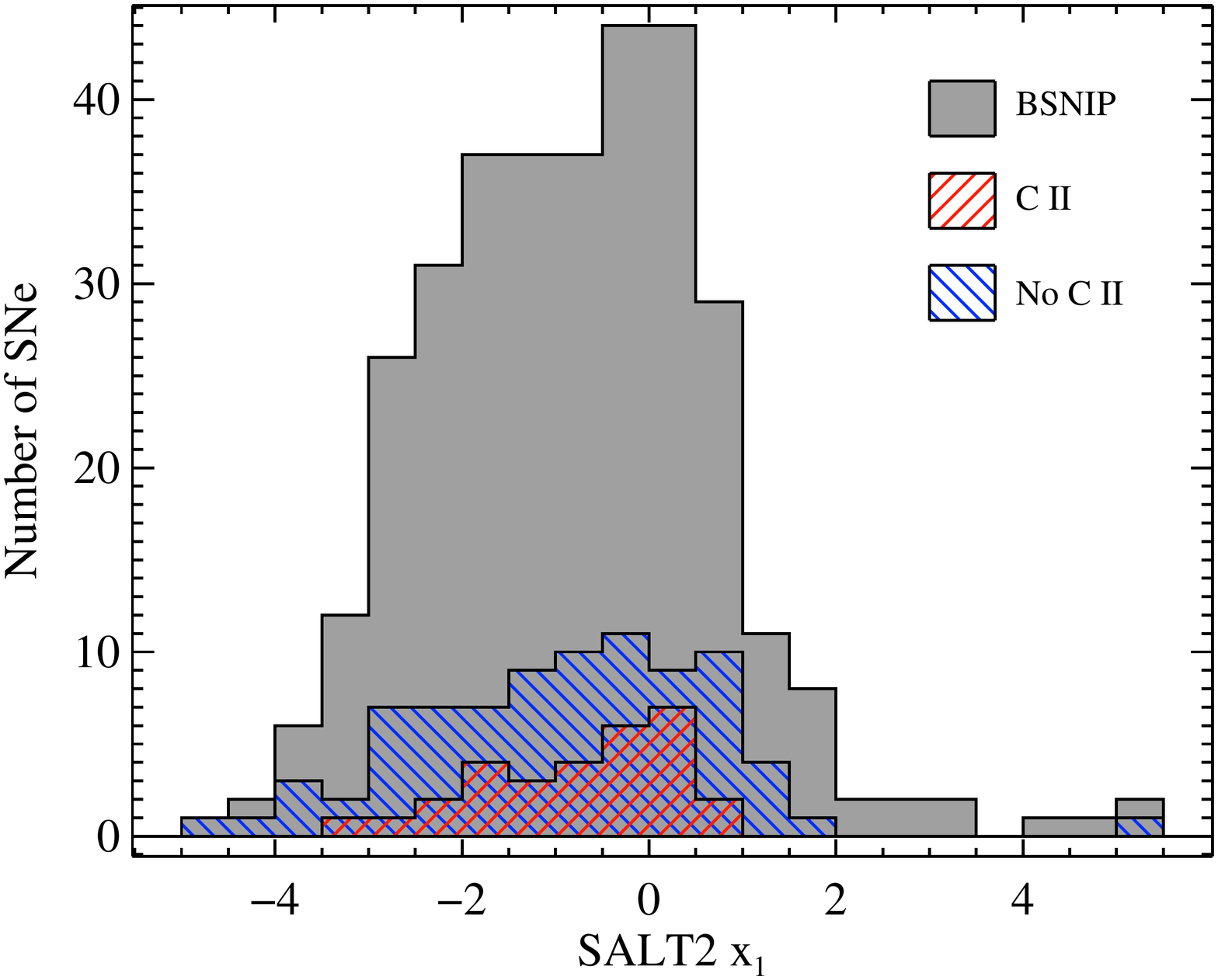} \\
\end{array}$
\caption{({\it Top}) A histogram of $\Delta m_{15}(B)$ for the
  entire BSNIP dataset (grey), with C (red hashed), and without C (blue
  hashed). There is no significant difference between $\Delta
  m_{15}(B)$ values for any of the three samples. ({\it Bottom}) A 
  histogram of $x_1$ using the same colours as the top panel. Again, there
  is no significant difference between $x_1$ values for any of the
  samples.}\label{f:width_hist} 
\end{figure}

As mentioned above, the Phillips relation correlates the peak
luminosity of a SN~Ia to its light-curve decline rate
\citep{Phillips93}. One way to 
parametrise this decline rate is to calculate the difference in
magnitudes between maximum and fifteen days past maximum in the
$B$ band, referred to as $\Delta m_{15}(B)$. The BSNIP sample
has 196 objects for which we calculate $\Delta m_{15}(B)$ and of
those, 28 show C and 67 show no C. A histogram of $\Delta m_{15}(B)$
values can be found in the top panel of Figure~\ref{f:width_hist}. The
average $\Delta m_{15}(B)$ for each of the three samples show in the
figure (all of
BSNIP, with C, and without C) are consistent with each other. This
was hinted at by \citet{Folatelli11}, though they admit that they have
too few objects to make any robust statistical statement. There also
appears to be no difference in $\Delta m_{15}(B)$ values when the
``with C'' sample is subdivided into `A' and `F' objects.

We also use an alternative parametrisation of the light-curve width:
the $x_1$ parameter from SALT2 \citep{Guy07}. The value of $x_1$ is in
the sense opposite that of $\Delta m_{15}(B)$. Thus, underluminous, narrow,
fast-evolving light curves have large values of $\Delta m_{15}(B)$ but
small values of $x_1$. There are 335 SNe~Ia in BSNIP that have a 
SALT2 fit, and 30 (83) of them exhibit (do not exhibit) C. A histogram
of $x_1$ values is shown in the bottom panel of
Figure~\ref{f:width_hist}.

Like $\Delta m_{15}(B)$, the average $x_1$
value for each sample is consistent with one another. This is at
odds with the finding of \citet{Thomas11} that SNe~Ia with C have
lower $x_1$ values when compared to SNe without C. Their C-positive
objects are mostly clustered near $x_1 \approx -2$ while our SNe~Ia
with C have a wide range of $x_1$ values, with the average being $x_1
\approx -0.74$ and the peak of the distribution occurring at $x_1
\approx 0$. However, the {\it overall} distributions of $x_1$ values
are different between \citet{Thomas11} and the current study; there
are a significant number of SNe~Ia with $x_1 < -2.5$ in the BSNIP
sample while there are none shown by \citet{Thomas11}.

\begin{figure}
\centering$
\begin{array}{c}
\includegraphics[width=3.4in]{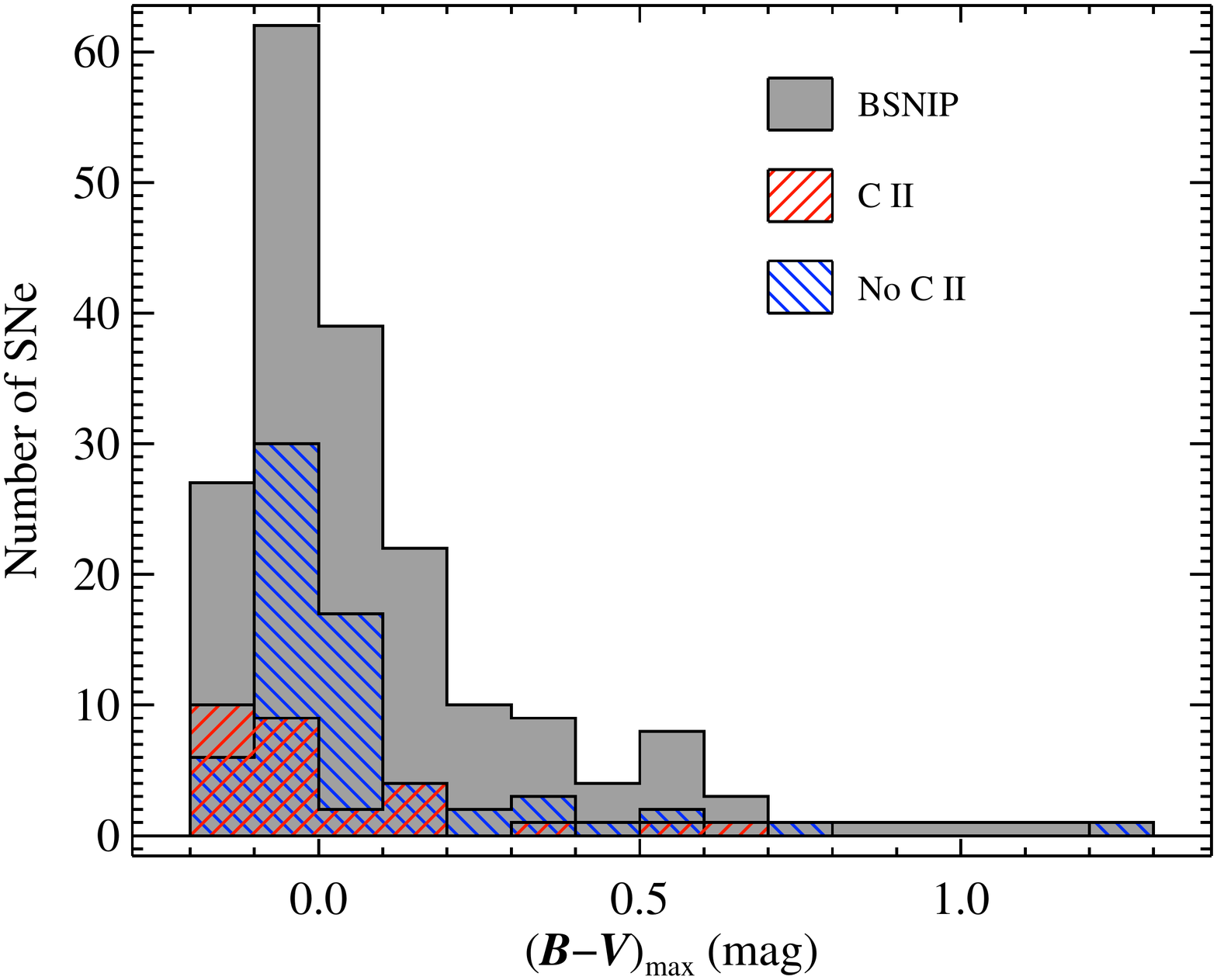} \\
\includegraphics[width=3.4in]{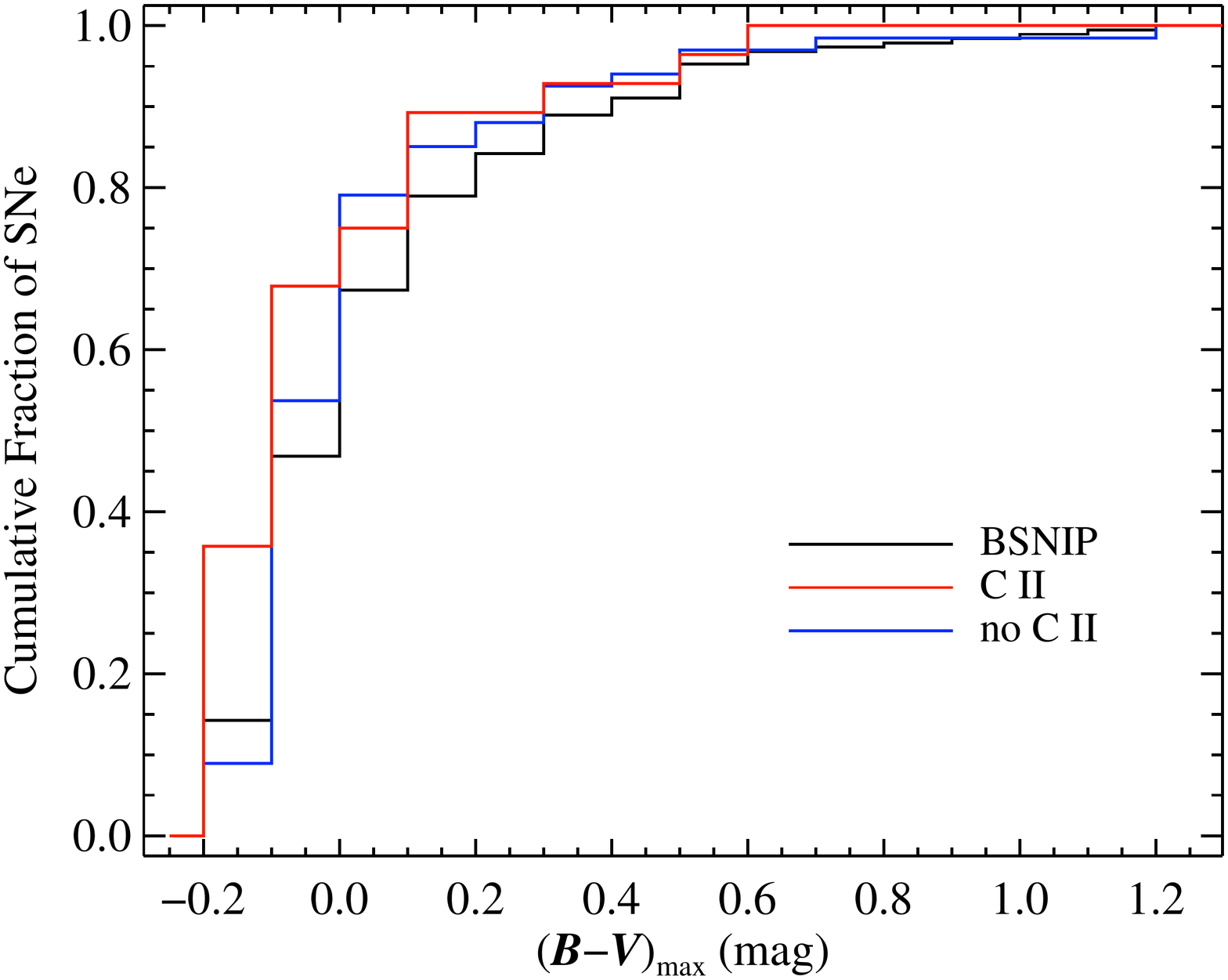} \\
\end{array}$
\caption{({\it Top}) A histogram of the difference between SN $B$-band 
  magnitude and $V$-band magnitude at the time of $B$-band maximum brightness
  (i.e., \bvmax) for the
  entire BSNIP dataset (grey), with C (red hashed), and without C (blue
  hashed). ({\it Bottom}) The cumulative distribution function of 
  \bvmax\ values using the same colours as the top panel. SNe~Ia with C
  tend to have bluer colours than those without C.}\label{f:bv} 
\end{figure}

The colours of SNe~Ia that show or do not show C signatures can also
be investigated. One way to quantify the colour of a SN~Ia is to
measure the difference between its $B$-band magnitude and $V$-band
magnitude at the time of $B$-band maximum brightness (referred to in
this work as \bvmax). The BSNIP data contain 190 objects for which
\bvmax\ is measured, 28 of which have C and 67 of which do not. While
there is no significant difference in \bvmax\ when the ``with C''
sample is subdivided into `A' and `F' objects, there {\it is} a
difference between the ``with C'' objects and the ``without C''
objects: SNe~Ia with C are bluer. A Kolmogorov-Smirnov (KS) test on
these two samples implies that they very likely come from different
parent populations ($p \approx 0.07$). The top panel of
Figure~\ref{f:bv} shows a histogram of \bvmax\ values for the entire
BSNIP dataset, objects with C, and those without. The bottom panel
shows the cumulative distribution function of these three samples. 

\begin{figure}
\centering$
\begin{array}{c}
\includegraphics[width=3.4in]{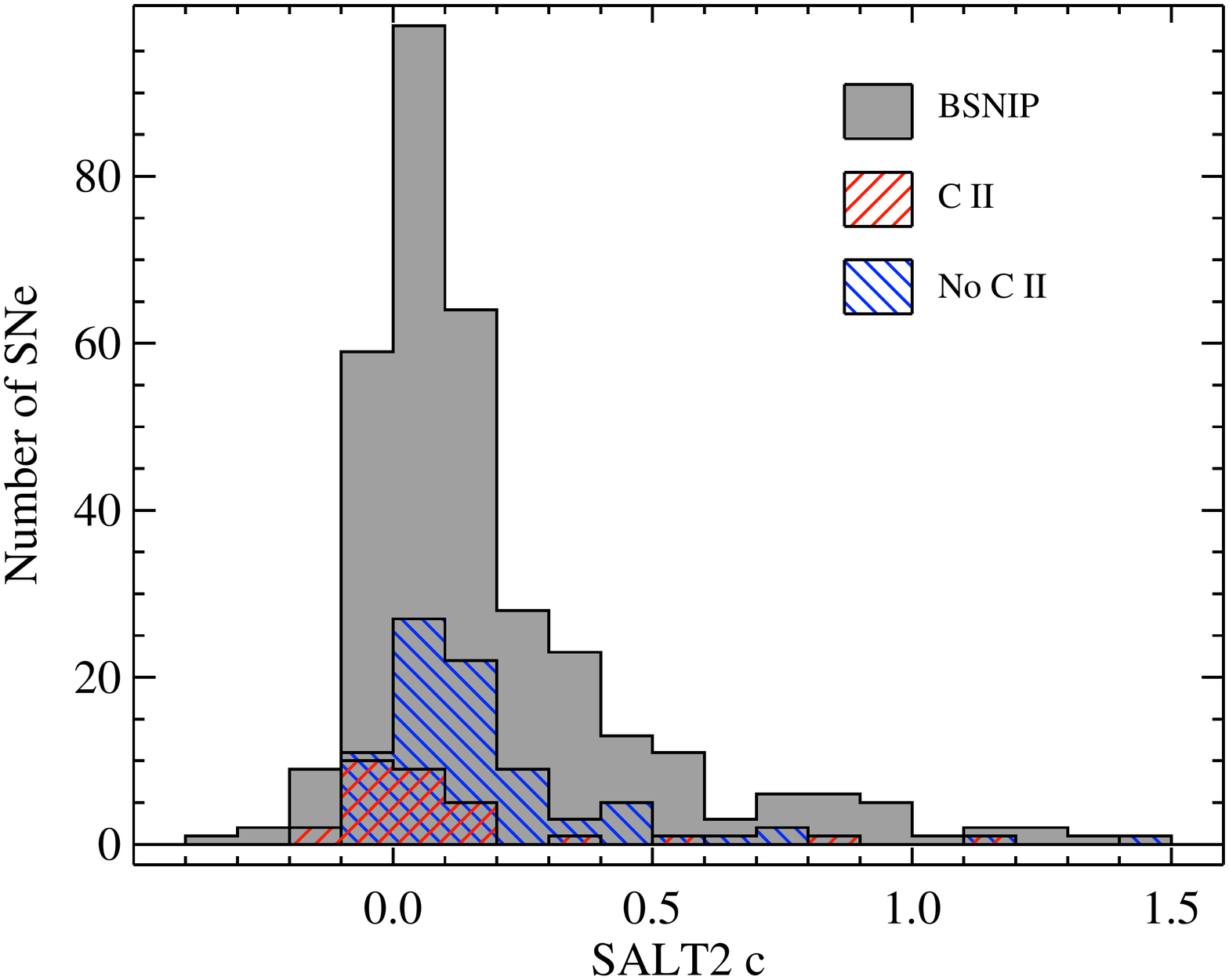} \\
\includegraphics[width=3.4in]{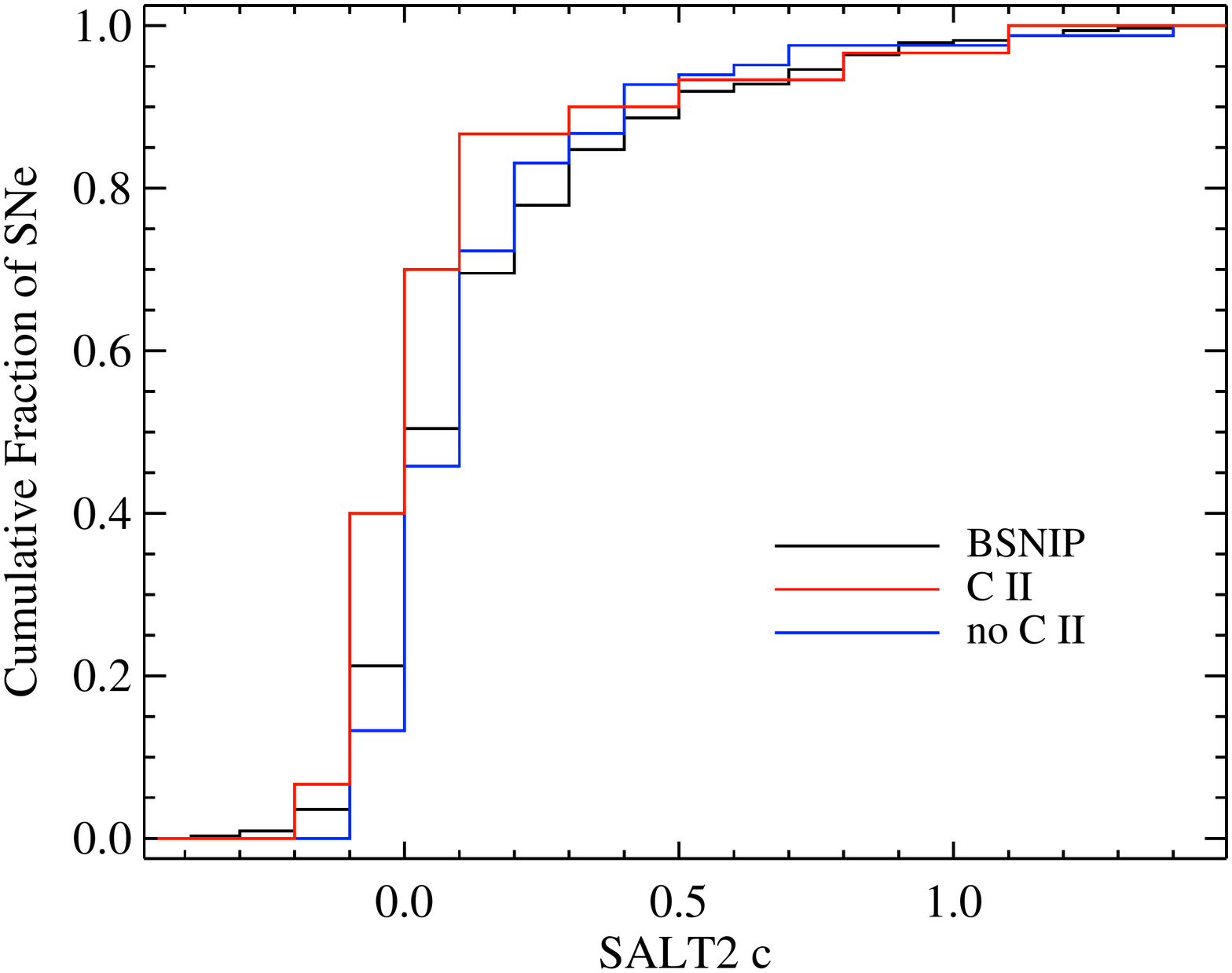} \\
\end{array}$
\caption{({\it Top}) A histogram of SALT2 $c$ for the
  entire BSNIP dataset (grey), with C (red hashed), and without C (blue
  hashed). ({\it Bottom}) The cumulative distribution function of SALT2 
  $c$ values using the same colours as the top panel. Again, SNe~Ia
  with C tend to have bluer colours than those without C.}\label{f:c} 
\end{figure}

This trend can also be seen if one parametrises SN colour by the
SALT2 $c$ parameter \citep{Guy07}. Of the 335 objects in BSNIP with
good SALT2 fits, 30 have C and 83 do not. And, as with the \bvmax\
values, C-positive objects appear to have bluer colours than C-negative
ones (a KS test yields $p \approx 0.01$), while no significant colour
difference is found between `A' objects and `F'
objects. Figure~\ref{f:c} presents the histogram (top panel) and the
cumulative distribution function (bottom panel) of the SALT2 $c$
values for each of the three samples.

Yet another way to quantify the colour of a SN~Ia is to calculate
synthetic photometric colours from a spectrum. In BSNIP~I it was shown 
that the relative spectrophotometry of our data, when compared to the
actual light curves, is accurate to $\le 0.07$~mag across the entire
spectrum. Therefore, synthetic colours derived from BSNIP spectra
should be photometrically accurate to about this level. In order to
determine the synthetic colours of our spectra 
in this work, we follow the procedure from BSNIP~I. Simply stated, we
convolve each spectrum with the \citet{Bessell90} filter functions,
which have approximate wavelength ranges of
3400--4100, 3700--5500, 4800--6900, 5600--8500, and 7100--9100~\AA\
for $U$, $B$, $V$, $R$, and $I$, respectively. We then calculate the
$U-B$, $B-V$, $V-R$, and $R-I$ colours. Most of the spectra
studied herein fully cover the \bvri bands and about half cover the
$U$ band as well. Using synthetic colours calculated from our spectra,
objects with C signatures once again have significantly bluer $U-B$
and $B-V$ colours at all epochs. On the other hand, for all other
colours calculated, the data are consistent with C-positive and
C-negative objects having similar colours.

The result found here that SNe~Ia with evidence for 
C tend to have bluer optical/near-ultraviolet (NUV) colours confirms
the work of previous groups 
\citep{Thomas11,Folatelli11}. A relationship between colour and
light-curve width was shown by \citet{Folatelli11} to be steeper for
objects with C as compared to objects without C when only SNe~Ia with
little intrinsic reddening were considered. The BSNIP data show no
significant evidence of different light-curve width versus colour
relationships for objects with or without C, whether we use all
objects or just those that are unreddened.

\citet{Milne10}, \citet{Thomas11}, and \citet{Milne12} present
{\it Swift}/UVOT \citep{Gehrels04,Roming05} 
photometry of a handful of SNe~Ia. In all of these works it is shown
that the objects that are relatively bright in the NUV
(``NUV blue''; i.e., those with the largest NUV
excesses) also show strong evidence for C absorption. However, they 
note that SN~2005cf, an object which is clearly C positive, has
completely normal colours in the {\it Swift}/UVOT data. All of the 
NUV-blue objects presented by \citet{Thomas11} and \citet{Milne12}
that are also studied in this work are found to exhibit C features, and
we find no evidence for C in three SNe~Ia that are NUV-red (according to
the {\it Swift}/UVOT data). However, two objects in BSNIP that are NUV-red
appear to have C signatures in their spectra: SN~2005cf (as mentioned
above) and SN~2007cq (the {\it Swift}/UVOT data indicate that it is NUV-red,
even though it has quite blue optical colours --- its \bvmax\ and
SALT2 $c$ are low, 0.004~mag and 0.024, respectively).

In summary, SNe~Ia which show evidence for C in their pre-maximum
spectra have bluer colours in the optical bands at all pre-maximum
epochs and are {\it almost} always found to be NUV-blue in space-based
UV/optical photometry. On the other hand, all SNe~Ia which are found
to be NUV-blue in the {\it Swift}/UVOT data are C-positive objects. Finally,
SNe~Ia which show no  evidence for C always have redder colours in
both the optical and NUV.

\subsection{\ion{C}{II} Velocities}\label{ss:vel}

\begin{figure}
\centering$
\begin{array}{c}
\includegraphics[width=3.4in]{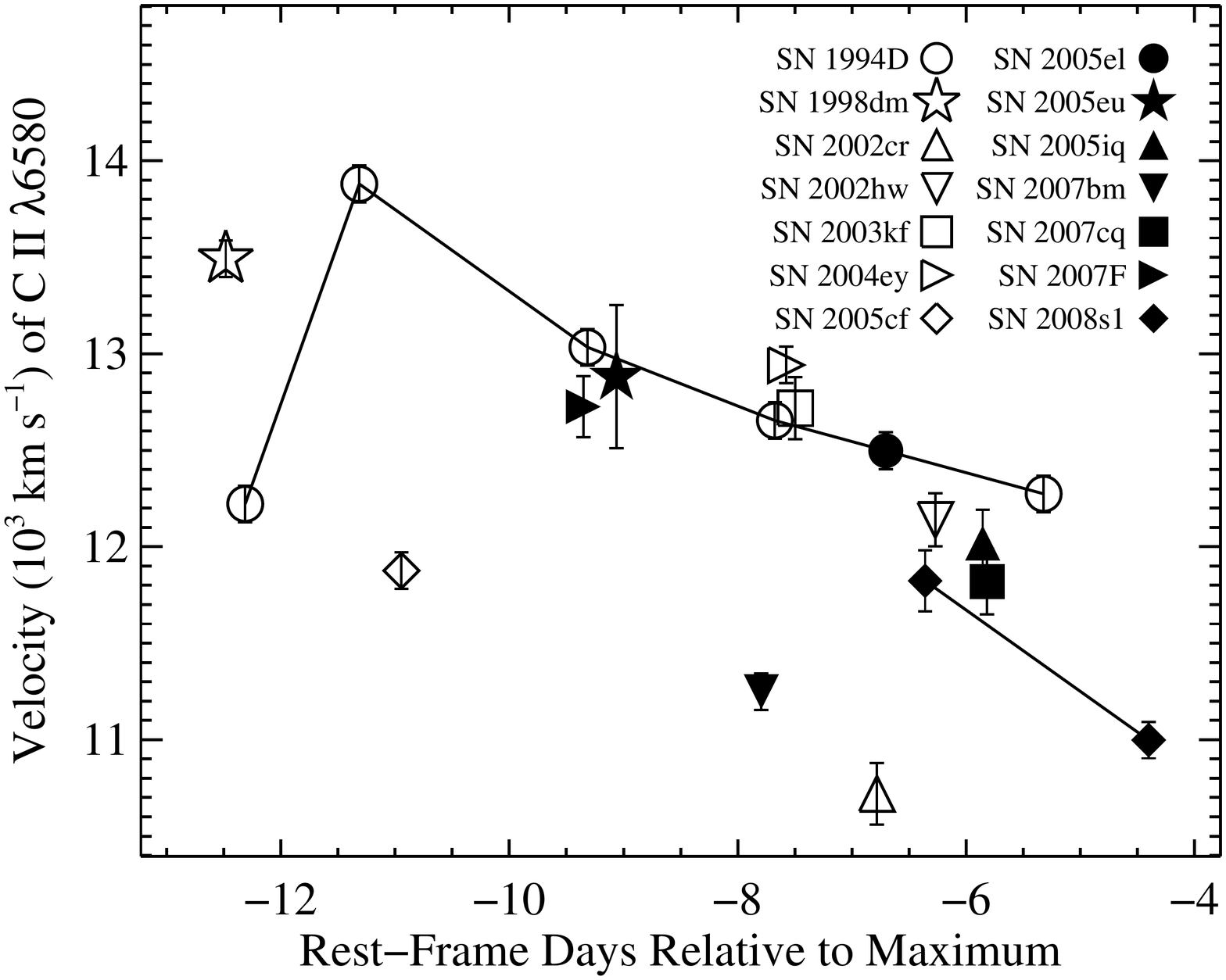} \\
\includegraphics[width=3.4in]{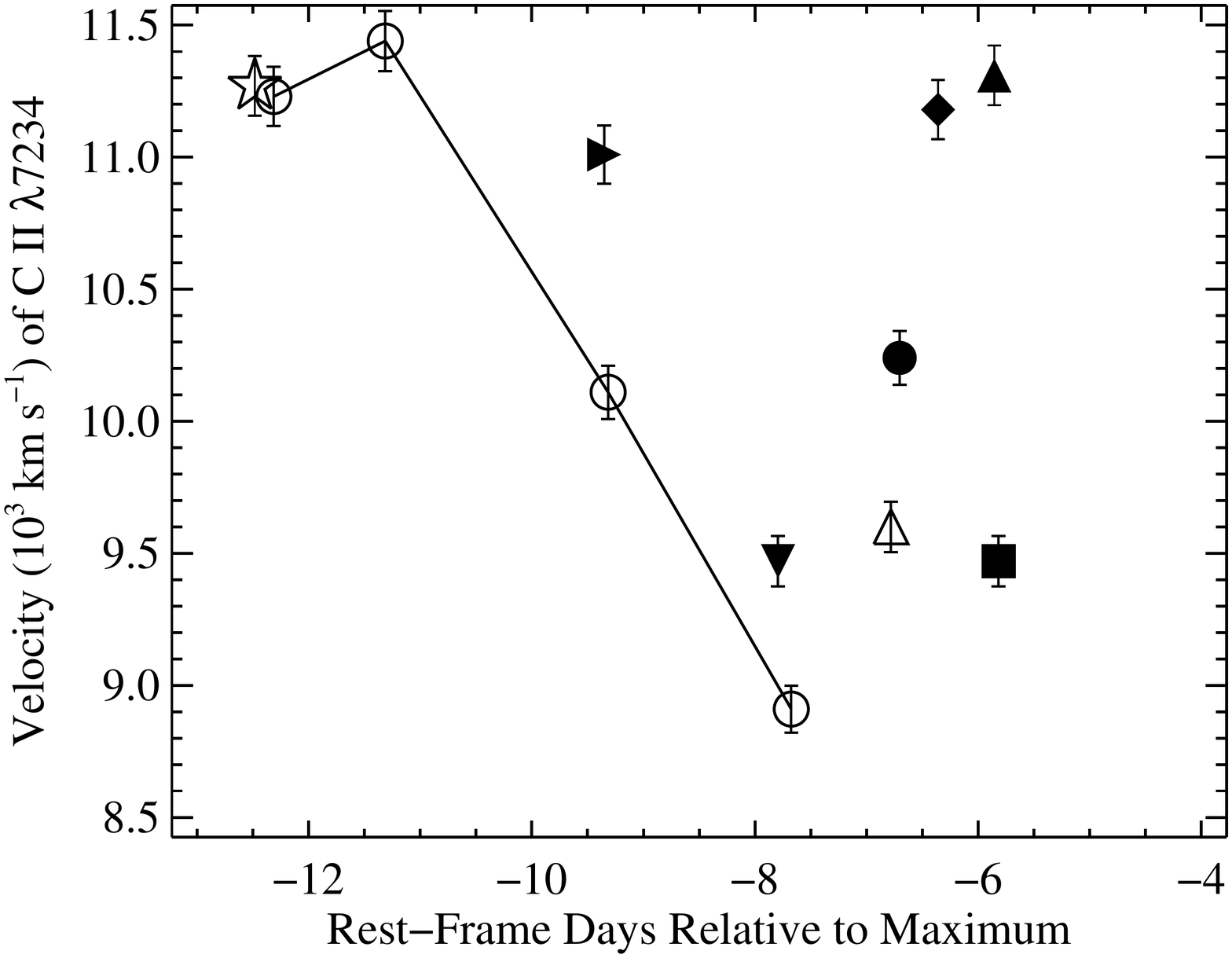} \\
\end{array}$
\caption{The temporal evolution of the expansion velocity of
  \ion{C}{II} $\lambda$6580 ({\it top}) and \ion{C}{II} $\lambda$7234
  ({\it bottom}). The different shapes correspond to different
  SNe. The two objects which have multiple velocity measurements have
  their velocities connected with a solid line.}\label{f:c_vel} 
\end{figure}

As described in Section~\ref{ss:measure}, we measure the expansion
velocity and pEW of the \ion{C}{II} $\lambda$6580 line in all 19
spectra classified as `A.' We also measure the velocity and pEW of the
\ion{C}{II} $\lambda$7234 line in 12 of these spectra. The temporal
evolution of the expansion velocities for both features can be found
in Figure~\ref{f:c_vel}. The different shapes correspond to different
SNe, and for the two objects which have multiple velocity measurements
(SN~1994D and SN~2008s1\footnote{Also known as SNF20080514-002.}),
their velocities are connected with a solid line.

The range of velocities spanned by both features is relatively
small. The \ion{C}{II} $\lambda$6580 line mostly has velocities
around 12,000--13,000~\kms, while the \ion{C}{II} $\lambda$7234 line
is mainly found between 9500~\kms\ and 11,500~\kms. However, this is a
somewhat larger range of $\lambda$6580 velocities than what has been
seen in previous work \citep{Folatelli11}. The largest \ion{C}{II}
velocity observed in the BSNIP data is \about14,000~\kms, which may be
caused more by an observational bias than a real, physical limit. Both
\citet{Parrent11} and \citet{Folatelli11} discuss the difficulty of
measuring \ion{C}{II} $\lambda$6580 with $v \ga 15,000$~\kms\ due to
the fact that at these high velocities the feature becomes strongly
blended with \ion{Si}{II} $\lambda$6355.

As seen in Figure~\ref{f:c_vel}, the typical \ion{C}{II} velocities of
all objects at a given epoch 
decrease with time \citep[as has been seen before;][]{Folatelli11},
though there is a dramatic {\it increase} in velocity between the
first and second epochs of SN~1994D. This has not been seen in
previous work, likely due to the fact that our data were obtained at
earlier epochs. Furthermore, \citet{Thomas11} measure all
\ion{C}{II} $\lambda$6580 velocities to be \about12,000~\kms, and they
see very little change with time.

The difference in the typical velocities of the two \ion{C}{II}
features implies that the  $\lambda$7234 line is \about2000~\kms\
slower than the $\lambda$6580 line.  There has been little attempt
previously to determine the velocity of \ion{C}{II}
$\lambda$7234 due to its relative weakness, but
\citet{Thomas11} do mention possible detections of this absorption at
velocities somewhat lower than those of \ion{C}{II} $\lambda$6580
(consistent with what is found here).

\begin{figure}
\centering$
\begin{array}{c}
\includegraphics[width=3.4in]{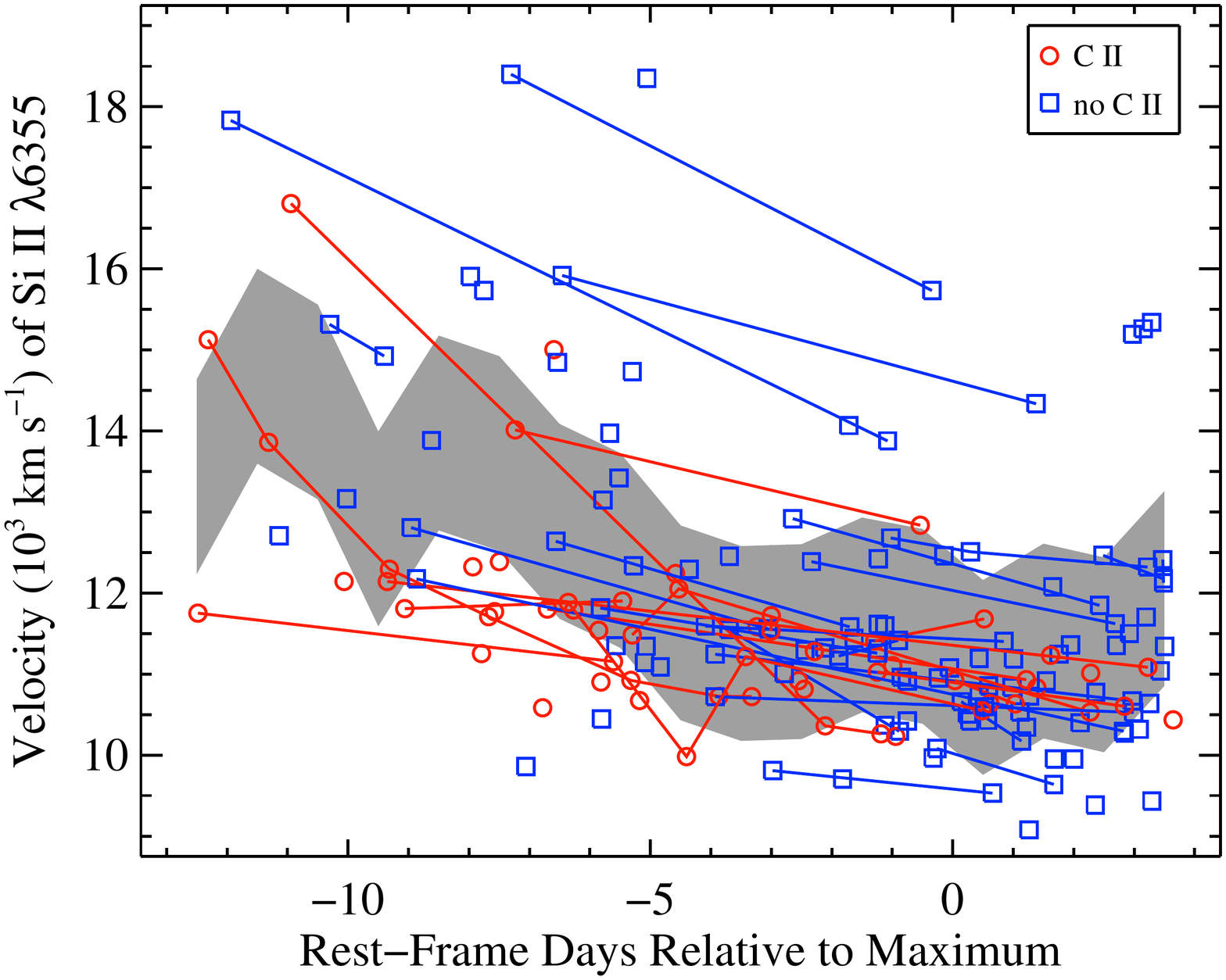} \\
\includegraphics[width=3.4in]{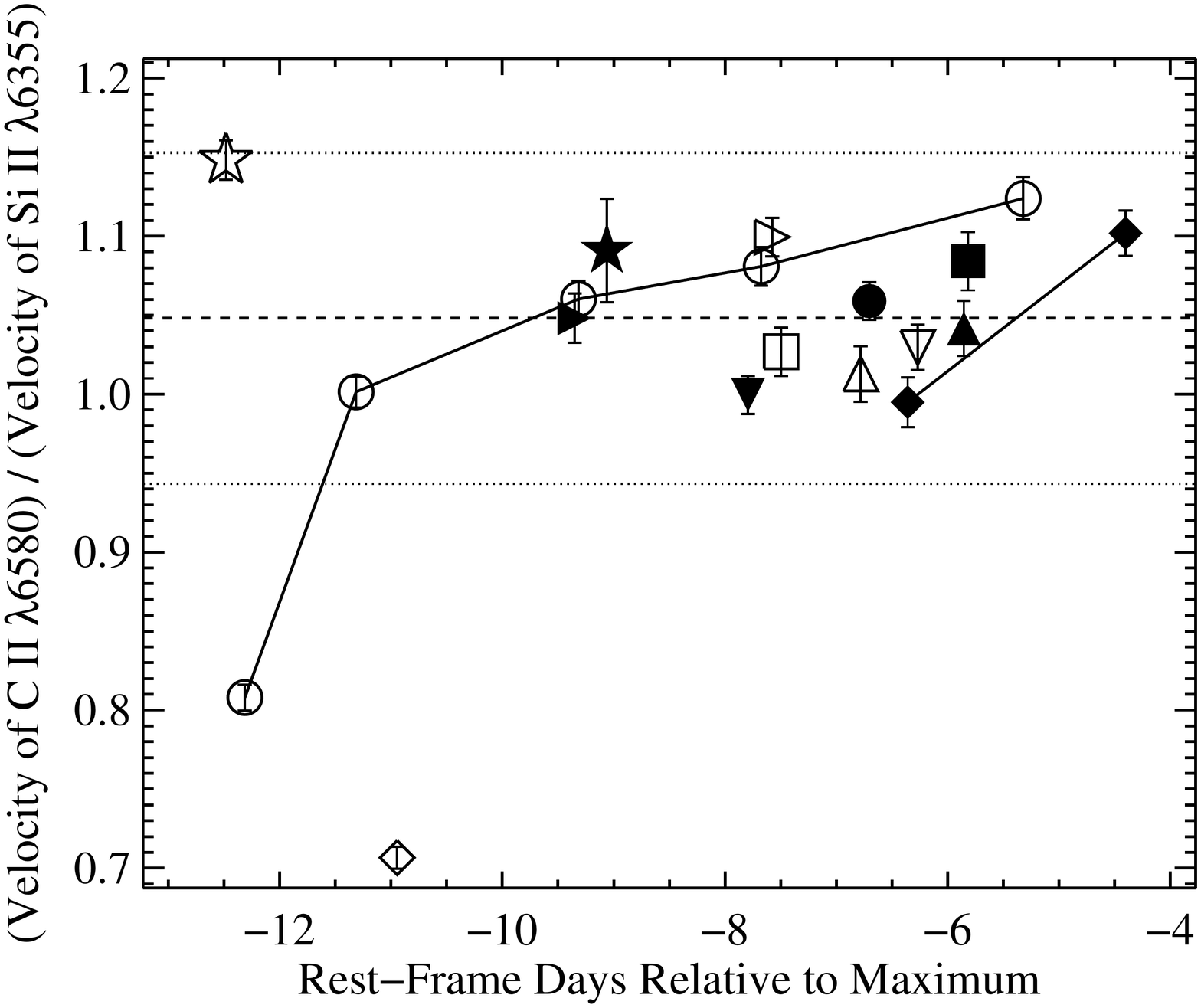} \\
\end{array}$
\caption{({\it Top}) The temporal evolution of the \ion{Si}{II} $\lambda$6355
  velocities for SNe~Ia with C (red circles), without C
  (blue squares), and the 1$\sigma$ region around the average velocity
  as determined by the entire BSNIP sample (grey area). ({\it Bottom}) The 
  temporal evolution of the ratio of the \ion{C}{II} $\lambda$6580 velocity to
  the \ion{Si}{II} $\lambda$6355 velocity. The plot
  symbols are the same as in Figure~\ref{f:c_vel}. The dashed line is
  the median ratio (\about1.05) and the dotted lines are the median
  $\pm 10$~per~cent. In both panels,  
  objects with multiple velocity measurements are connected with a
  solid line.}\label{f:si_vel} 
\end{figure}

The \ion{Si}{II} $\lambda$6355 velocities (as measured in BSNIP~II) of
the objects with and without C can also be compared and are shown in
the top panel of Figure~\ref{f:si_vel}. Red circles are SNe~Ia with C
and blue squares do not have C. The grey shaded area represents the
1$\sigma$ region around the average \ion{Si}{II} $\lambda$6355
velocity from the entire BSNIP sample. Individual objects with
multiple velocity measurements are connected with a solid line. The
panel includes all 131 objects in this work with a definitive C
classification (`A,' `F,' or `N').

According to the figure, SNe~Ia with C tend to have lower
\ion{Si}{II} $\lambda$6355 velocities. Nearly all of 
the C-positive spectra have \ion{Si}{II} velocities that are at or
below average, while the objects without C span the entire range of
\ion{Si}{II} velocities from below average to well above average. This
lack of HV objects that show C was mentioned above, and is likely in
part due
to the difficulty in detecting and measuring \ion{C}{II} $\lambda$6580
at large expansion velocities. In fact, \citet{Folatelli11} found that
for all C-positive objects in their sample, the \ion{Si}{II}
$\lambda$6355 velocities were $<$12,500~\kms. We find six
spectra of C-positive SNe~Ia that have \ion{Si}{II} velocities above
this value, and three of them are from epochs earlier than the earliest
ones studied by \citet{Folatelli11}, when one expects even larger
expansion velocities for all elements. Thus, our findings appear to be
consistent with those of previous work that the \ion{Si}{II}
$\lambda$6355 velocities of C-positive SNe~Ia are significantly {\it
  lower} than average.

The typical \ion{Si}{II} $\lambda$6355 velocities for the C-positive
objects are 10,000--12,000~\kms, very close to the typical
\ion{C}{II} $\lambda$7234 velocities. However, this is 1000--2000~\kms\
{\it slower} than the typical \ion{C}{II} $\lambda$6580 velocities, as
was also found previously \citep{Folatelli11}. The bottom panel of
Figure~\ref{f:si_vel} shows the ratio of the \ion{C}{II} $\lambda$6580
velocity to the \ion{Si}{II} $\lambda$6355 velocity for all 19 `A'
spectra. The plot symbols are the same as in Figure~\ref{f:c_vel}, and
again the two objects having multiple velocity measurements are
connected with a solid line. The dashed line is the median ratio
(\about1.05) and the dotted lines are the median $\pm 10$~per~cent.

The ratio of these two velocities is remarkably constant, especially
for $t > -10$~d. A similar trend was found by \citet{Parrent11},
though their average ratio was slightly larger than ours (1.1). For a
given object, the ratio may increase somewhat with time, but with only
two objects with multiple velocity measurements in our sample it is
difficult to make any definitive statement about this. However, the
data presented by \citet{Parrent11} seem to support this conclusion as
well. Furthermore, we note that the \ion{C}{II} velocities are {\it
  usually} similar to or larger than the \ion{Si}{II} velocities,
which supports the idea of the layered structure of SN~Ia ejecta with
some additional mixing. The standard layering picture includes
unburned C in layers that are further out (i.e., faster expanding)
than the layers containing newly synthesised Si. However, some degree
of mixing between these layers is required in order to reproduce the
observed overlap in velocity space of the \ion{C}{II} and \ion{Si}{II}
features.

For the spectra with $t < -10$~d, the two outliers on the low end (the
first epoch of SN~1994D and SN~2005cf) both have relatively low
\ion{C}{II} $\lambda$6580 velocities {\it and} higher than average
\ion{Si}{II} $\lambda$6355 velocities (leading to a small
ratio). \citet{Parrent11} found no normal SNe~Ia with a ratio much
less than 1, but their ratio for SN~1994D at $t \approx -11$~d is
\about1, which matches very well our second epoch of SN~1994D. Thus,
these abnormally low velocity ratios at early epochs may be real and
should be investigated further in the future with more early-time spectra.
SN~1998dm appears to be somewhat of an outlier ($<$10~per~cent
above the median ratio) at the high end at early times; it has a higher 
than normal \ion{C}{II} $\lambda$6580 velocity with a relatively 
normal \ion{Si}{II} $\lambda$6355 velocity (yielding a larger ratio).

Comparing C features to O features (specifically, the \ion{O}{I}
triplet centered near $\lambda$7773) may also be interesting since O
is found in SN~Ia ejecta as unburned fuel and a product of C
burning. While the discussion of how to distinguish between O that is
fuel and O that is ash is beyond the 
scope of this paper, we nonetheless compare our \ion{C}{II}
measurements to \ion{O}{I} triplet measurements taken from
BSNIP~II. However, we note that the \ion{O}{I} triplet is 
notoriously difficult to measure accurately due to the fact that it is
highly contaminated by telluric absorption features. Moreover, the
\ion{O}{I} triplet is quite weak at the early epochs studied
herein.

Those caveats notwithstanding, we find that objects with and
without evidence for \ion{C}{II} have similar \ion{O}{I} triplet
velocitie. Furthermore, the \ion{O}{I} velocities of C-positive
objects closely follow the average \ion{O}{I} velocities of the entire
BSNIP sample. Finally, there are two spectra (of two objects) for
which we measure velocities of {\it both} \ion{C}{II} $\lambda$6580
and the \ion{O}{I} triplet, and we find that the velocities of these
two features are effectively equal to each other in a given spectrum.

We also investigated any possible correlations between \ion{C}{II}
velocities and photometric parameters. No significant
correlations were found between \ion{C}{II} velocity and light-curve
width (parametrised by $\Delta m_{15}(B)$ or SALT2 $x_1$) or SN colour
(parametrised by \bvmax\ or SALT2 $c$).

\subsection{\ion{C}{II} pEWs}\label{ss:ew}

\begin{figure}
\centering$
\begin{array}{c}
\includegraphics[width=3.4in]{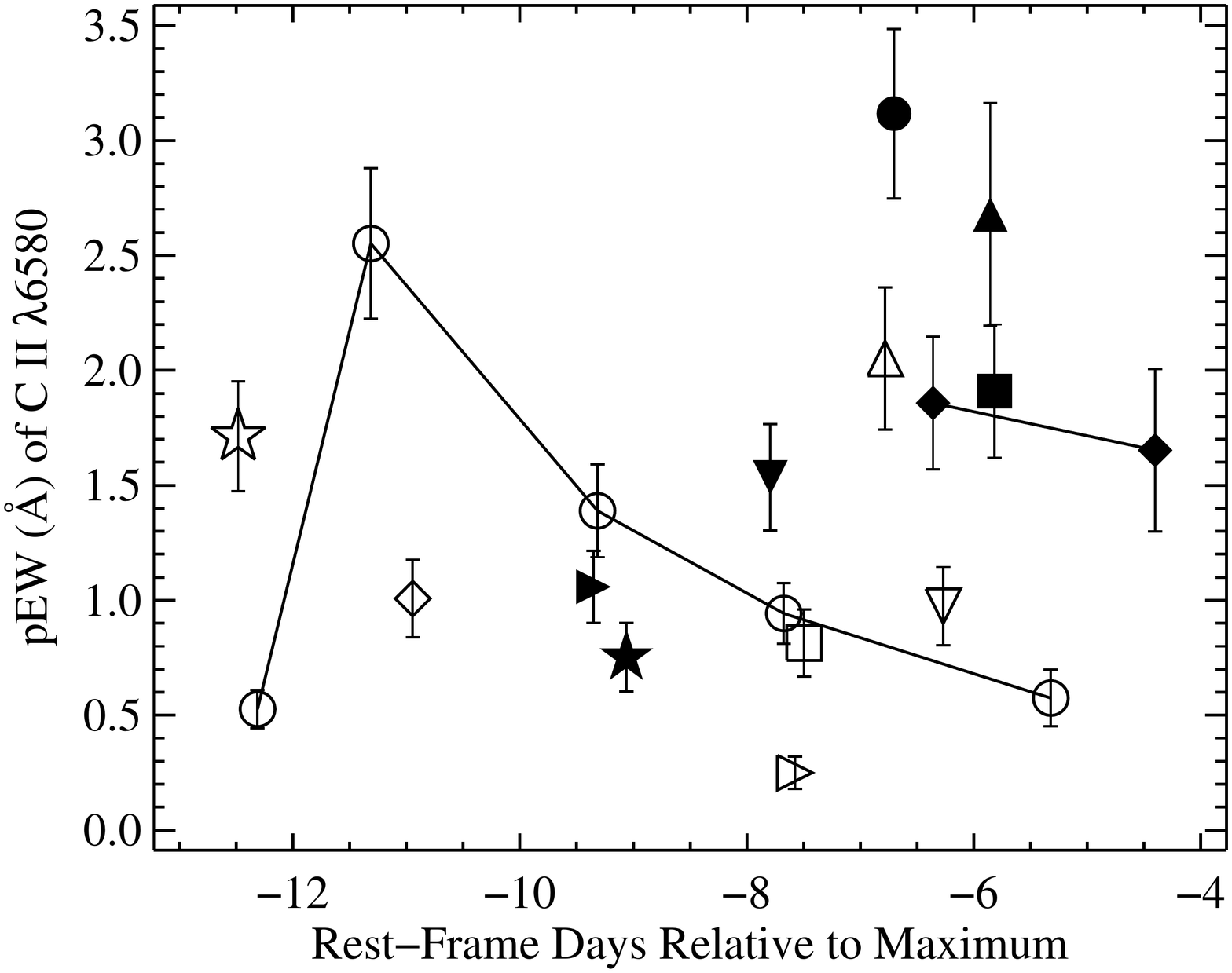} \\
\includegraphics[width=3.4in]{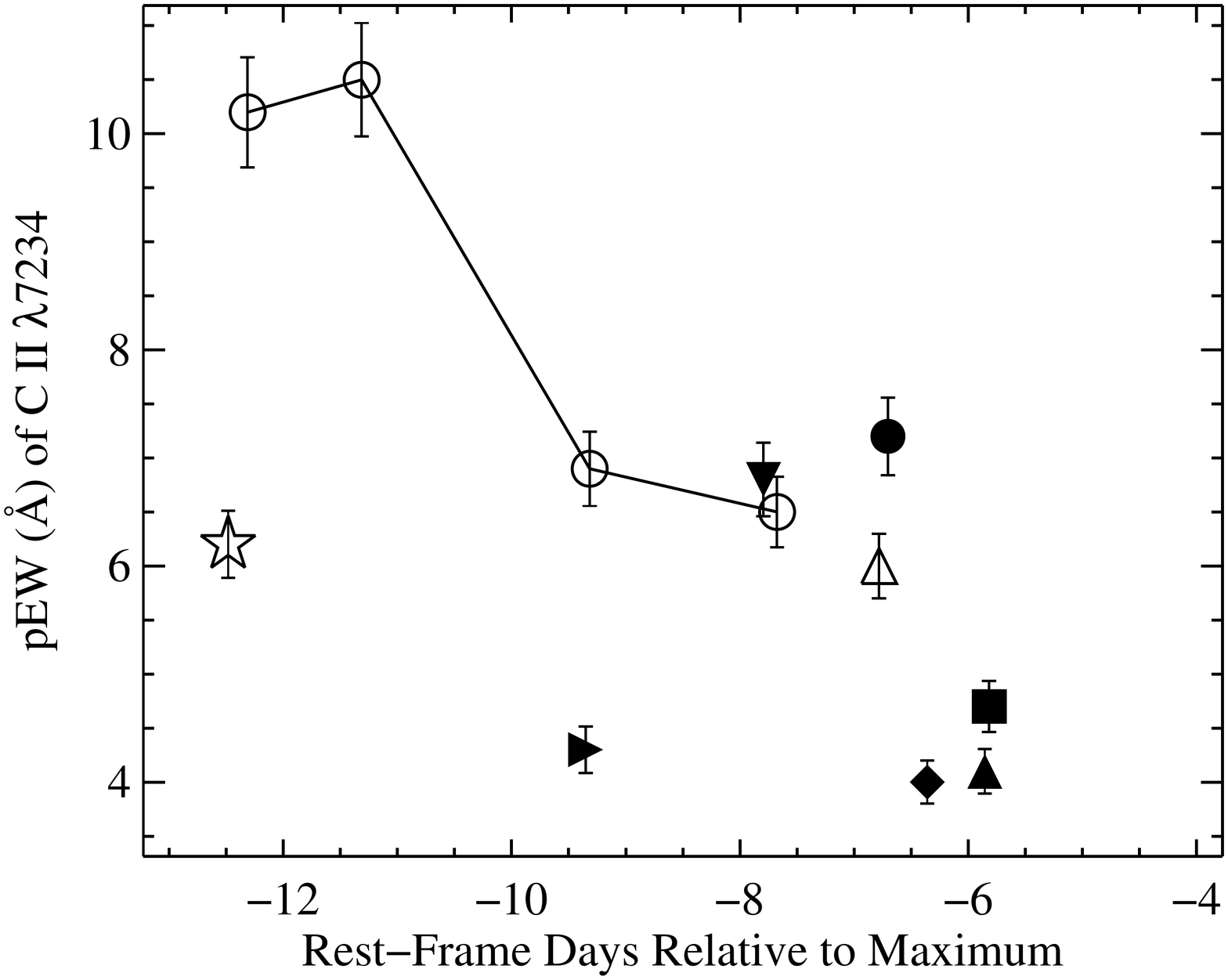} \\
\end{array}$
\caption{The temporal evolution of the pEW of
  \ion{C}{II} $\lambda$6580 ({\it top}) and \ion{C}{II} $\lambda$7234
  ({\it bottom}). The plot symbols are the same as in
  Figure~\ref{f:c_vel}. In both panels, 
  objects with multiple pEW measurements are connected with a
  solid line.}\label{f:c_ew} 
\end{figure}

The temporal evolution of the pEWs for both features can be found in
Figure~\ref{f:c_ew}. The plot symbols are the same as in
Figure~\ref{f:c_vel}, and the two objects which have multiple pEW
measurements are connected with a solid line. The \ion{C}{II}
$\lambda$6580 feature has pEWs which are all $<$3.5~\AA,
consistent with what has been seen previously \citep{Folatelli11}. The
\ion{C}{II} $\lambda$7234 feature, on the other hand, has not been
measured before, and we find a range of pEW values of
\about4--11~\AA.

As seen in Section~\ref{ss:time}, the probability of detecting C
decreases with time, mostly due to a weakening of the \ion{C}{II}
absorption features. Figure~\ref{f:c_ew} shows evidence to support
the idea that, for the most part, the pEWs of the two \ion{C}{II}
features decrease with time. However, we must point out that there are
only two objects with multiple pEW measurements. Interestingly, there
appears to be an increase in the pEW for one of these objects (SN~1994D)
at the earliest epochs ($-13\textrm{ d} \la t \la -11$~d). While the
increase is marginal and consistent with no change in pEW for the
\ion{C}{II} $\lambda$7234 feature, the pEW increase is quite
significant for the \ion{C}{II} $\lambda$6580 feature (this can be
seen visually in the top-left panel of Figure~\ref{f:carbons}).

It was suggested by \citet{Folatelli11} that one might observe such an 
increase in pEW between 13 and 11~d before maximum brightness, 
but their data did not extend to sufficiently early epochs 
to investigate this further. The expected increase in pEW was based on
synthetic spectra created using a Monte Carlo code
\citep{Mazzali93,Lucy99,Mazzali00} as implemented in the analysis of
SN~2003du \citep{Tanaka11}. \citet{Folatelli11} find that the
synthetic spectra at 
these early epochs show a strong, red emission component of the
\ion{Si}{II} $\lambda$6355 feature which tends to ``fill in'' some of
the \ion{C}{II} $\lambda$6580 absorption, thus leading to a low pEW
measurement for C. Furthermore, they point out that at these epochs
some C is below the photosphere (leading to a lower measured 
pEW), and that the large expansion velocities at these times make 
\ion{C}{II} $\lambda$6580 become increasingly blended with \ion{Si}{II}
$\lambda$6355 (again leading to difficulty in measuring pEWs of
\ion{C}{II}).

The model used for SN~2003du by \citet{Tanaka11} required $6.8 \times
10^{-3}$~\msun\ of C (mass fraction $X\left(C\right) = 0.002$) in the
velocity range $10,500 < v < 15,000$~\kms, and the pEWs measured from
these synthetic spectra are plotted as open red squares in Figure~9 of
\citet{Folatelli11}. This velocity range is consistent with the
velocities we find for the \ion{C}{II} $\lambda$6580 feature;
more impressively, the theoretical pEW values shown by
\citet{Folatelli11} are excellent matches to the 
pEWs we measure for SN~1994D. Furthermore, \citet{Folatelli11} discuss
two other models where the amount of C is increased and decreased by a
factor of four from the SN~2003du value. They state that this 
range of C mass ($1.7 \times 10^{-3}$ -- $2.7 \times 10^{-2}$~\msun)
includes objects where no C is detected as well as objects that have the
largest \ion{C}{II} $\lambda$6580 pEWs (at \about1 week before
maximum light). The pEWs measured from the BSNIP data span a similar range
of values as the data studied by \citet{Folatelli11}, and so we find
that this mass range for C also encompasses all of our data.

\begin{figure}
\centering
\includegraphics[width=3.4in]{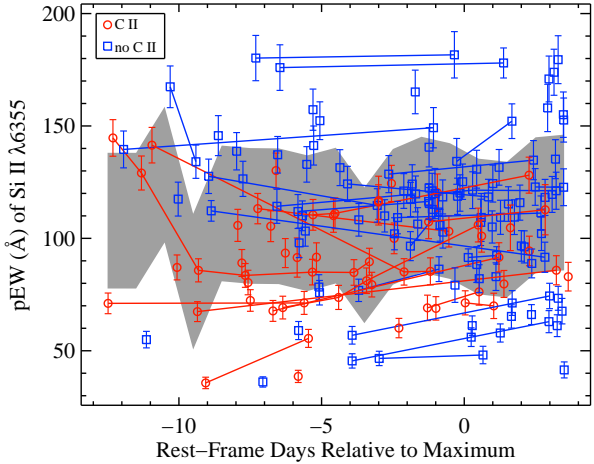}
\caption{The temporal evolution of the \ion{Si}{II} $\lambda$6355
  pEW for SNe~Ia with C (red circles), without C
  (blue squares), and the 1$\sigma$ region around the average velocity
  as determined by the entire BSNIP sample (grey area). 
  Objects with multiple velocity measurements are connected with a
  solid line.}\label{f:si_ew} 
\end{figure}

We plot the temporal evolution of the \ion{Si}{II} $\lambda$6355 pEWs
(as measured in BSNIP~II) of SNe~Ia with and without C in
Figure~\ref{f:si_ew}. As before, red circles are SNe~Ia with C, blue
squares do not have C, and the grey area is the 1$\sigma$ region
around the average \ion{Si}{II} $\lambda$6355 pEW from the entire
BSNIP sample. Again, individual objects with multiple measurements are
connected with a solid line.
Objects which show evidence for C tend to have slightly below average
\ion{Si}{II} $\lambda$6355 pEWs, while objects without C follow the
average pEW distribution quite well. However, the significance of this 
difference between C-positive objects and C-negative objects (or
the entire BSNIP sample) is relatively weak. This may yet again be the
observational bias that as the \ion{Si}{II} $\lambda$6355 pEW
increases, it becomes more blended with the \ion{C}{II} $\lambda$6580
feature (making C detection more difficult).

As with the \ion{C}{II} velocities, no significant correlations were
found between \ion{C}{II} pEW and light-curve width (parametrised by
$\Delta m_{15}(B)$ or SALT2 $x_1$) or SN colour (parametrised by
\bvmax\ or SALT2 $c$). Furthermore, no correlations were seen between
pEW and synthetic photometric colours as derived from the spectra
themselves (Section~\ref{ss:comparison}). Various spectroscopic
luminosity and colour indicators (which are defined and discussed at
length in BSNIP~II and BSNIP~III) were also found to be uncorrelated
with \ion{C}{II} pEW. Moreover, no relationship was found between the
pEW and velocity of either \ion{C}{II} feature.

\begin{figure}
\centering
\includegraphics[width=3.4in]{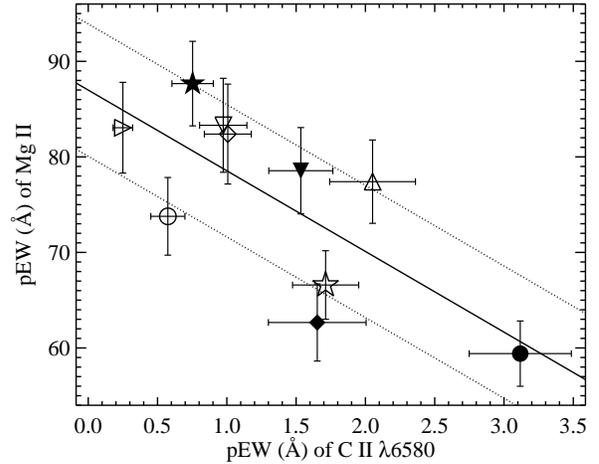}
\caption{The pEW of the \ion{C}{II} $\lambda$6580 feature versus the
  pEW of the \ion{Mg}{II} complex. The data are highly correlated with
  a correlation coefficient of $-0.73$. The solid line is the best
  linear fit to the data and the dotted lines are the root-mean square
  error. The plot 
  symbols are the same as in Figure~\ref{f:c_vel}.}\label{f:ewc_ewmg} 
\end{figure}

Objects with and without C have similar \ion{O}{I} triplet pEWs and
both samples are similar to the \ion{O}{I} triplet pEW distribution of
the full BSNIP sample. The pEWs of most other spectral features seen
in near-maximum spectra of SNe~Ia showed no significant correlation
with \ion{C}{II} pEW, except for the so-called \ion{Mg}{II} complex
(with a correlation coefficient of $-0.73$). Figure~\ref{f:ewc_ewmg}
shows the 10 objects which have measured pEW values for both
\ion{C}{II} $\lambda$6580 and the \ion{Mg}{II} complex. The solid line
is the best linear fit to the data and the dotted lines are the
root-mean square error. The plot symbols are the same as in 
Figure~\ref{f:c_vel}. If the two pEWs are measured in more than one
spectrum of a given object, we only plot the spectrum that is closest
to maximum brightness (i.e., the oldest) in Figure~\ref{f:ewc_ewmg}.

The \ion{Mg}{II} complex was defined in BSNIP~II and consists of a
blend of many IGE spectral lines which encompasses a broad, complex
absorption feature at 4100--4500~\AA. In BSNIP~III it was shown that
both the pEWs of the \ion{Mg}{II} and \ion{Fe}{II} (another broad
complex of IGE features at 4500--5200~\AA) complexes were relatively
good proxies for SALT2 colour \citep[$c$,][]{Guy07}, with increased
pEW implying a redder 
colour. While the \ion{Fe}{II} complex is not very well correlated
with pEW of \ion{C}{II} $\lambda$6580 (correlation coefficient of
$-0.32$), the strong anti-correlation between the pEW of the
\ion{Mg}{II} complex and the pEW of \ion{C}{II} indicates that {\it
  increased} pEW of \ion{C}{II} implies a {\it bluer} colour.

It has already been shown in this work that C-positive SNe~Ia tend to
have bluer colours than objects without C
(Section~\ref{ss:comparison}). In light of the relationship between
the pEWs of \ion{C}{II} and the \ion{Mg}{II} complex,
perhaps this relationship with colour extends further than a simple
binary splitting of objects with C versus those without C. It seems
possible that the strength of the C feature is directly related to
the colour of the SN. Objects where no C is detectable have the
reddest colours, while objects with some C (i.e., low pEWs) have
moderate colours, and finally objects with the most C (i.e., high
pEWs) have the bluest colours. Unfortunately, as mentioned above, the
pEW of the \ion{C}{II} $\lambda$6580 feature is not significantly
correlated with any direct measure of the SN colour (via light curves
or synthetic photometry from the spectra). The fact that the
\ion{C}{II} pEW appears well correlated with two pEW-based spectral
indicators of colour is intriguing, though, and should most definitely
be investigated further in future studies.


\section{Conclusions}\label{s:conclusions}

In this work we have searched for signatures of unburned C from the
progenitor WD using a subset of the BSNIP spectroscopic sample. We
classify 188 spectra of 144 SNe~Ia with ages $\la$3.6~d after maximum
brightness as either showing definite \ion{C}{II} absorption (`A'),
possibly showing evidence for C (`F'), definitely not showing C (`N'),
or inconclusive (`?'). The spectrum-synthesis code {\tt SYNOW} was
used to accurately classify all spectra that showed possible evidence
for C. The primary evidence for C is a distinct absorption line
associated with \ion{C}{II} $\lambda$6580, though absorption from
\ion{C}{II} $\lambda$7234 is also sometimes detected.

We find that \about11~per~cent of the SNe studied show definite C
absorption features, while a total of \about25~per~cent show at least
some evidence for C in their spectra, consistent with previous work
\citep{Parrent11,Thomas11,Folatelli11}.  The detection rate of C
decreases with time, though C can sometimes be seen at all ages younger 
than \about4~d past maximum brightness. Near 4~d {\it before} maximum
brightness, there is a 50~per~cent probability of detecting C,
according to the BSNIP data. If one obtains a spectrum at $t \la
-5$~d, then there is a better than 30~per~cent chance of detecting
a distinct absorption feature from \ion{C}{II}.

Nearly all objects that show C are spectroscopically normal (as
defined by various classification schemes), while SNe~Ia without C
detections are from all spectroscopic subtypes. The velocity
gradients of objects with and without C have a similar average and
range, and C detections and velocity gradients do not seem to be
related in any way. However, we again point out that the
BSNIP dataset is not well suited to velocity-gradient
measurements. The light curves of SNe~Ia with and without C also
appear to have the same distribution. On the other hand, confirming
previous work \citep{Thomas11,Folatelli11}, objects 
with C tend to have bluer optical colours than those without, and some
(but not all) also have strong NUV excesses. This is shown with the
BSNIP data using a variety of optical colour measurements.

The typical expansion velocity of the \ion{C}{II} $\lambda$6580
feature is 12,000--13,000~\kms, which is somewhat faster than the
usual velocity measured for the \ion{C}{II} $\lambda$7234 feature
(and we are the first to carefully study the velocity of this
feature). The \ion{Si}{II} $\lambda$6355 velocities, measured in
BSNIP~II, tend to be lower than average for C-positive objects, while
SNe~Ia without C have a wide range of \ion{Si}{II}
velocities. The ratio of the \ion{C}{II} $\lambda$6580 to
\ion{Si}{II} $\lambda$6355 velocities is remarkably constant with
time and among different objects, with a median value of \about1.05
\citep[consistent with what was reported by][]{Parrent11}.

The pEWs of the \ion{C}{II} $\lambda$6580 and \ion{C}{II}
$\lambda$7234 features are found mostly to decrease with time, though
there is a significant increase between \about13 and 11~d before
maximum light. This is consistent with the predictions made by
\citet{Folatelli11} from spectral models based on those presented by
\citet{Tanaka11}. The range of pEWs measured from the BSNIP data is
consistent with earlier work and implies a range of C mass in SN~Ia 
ejecta of $2 \times 10^{-3}$ -- $3 \times 10^{-2}$~\msun\
\citep{Folatelli11}. C-positive objects tend to have slightly lower
than average \ion{Si}{II} $\lambda$6355 pEWs, but at a relatively low
significance. The pEW of the \ion{Mg}{II} complex is found to be
strongly anti-correlated with the pEW of \ion{C}{II}
$\lambda$6580, implying that bluer objects should have larger
\ion{C}{II} pEWs. This is consistent withour finding that objects
with obvious C tend to have bluer optical colours than those
without. However, we find no strong correlation when comparing
the pEW of \ion{C}{II} to direct measures of SN colour. 

Even though this is the largest set of SNe~Ia for which C has ever
been searched, there are still only a handful of strong C
detections and measurements of C absorption features. Other studies
using independent datasets \citep{Thomas11,Folatelli11} and literature
searches \citep{Parrent11} have also been conducted, and we confirm
most of their findings at higher significance. Still, many more
moderate-to-high S/N SN~Ia spectra at early epochs are needed to
better investigate C and further probe WD progenitor models and SN~Ia
explosion mechanisms. New, large-scale transient searches such as
Pan-STARRS \citep{Kaiser02} and the Palomar Transient Factory
\citep[PTF;][]{Rau09,Law09} will be critical to moving this topic forward
as they find progressively more young SNe of all types. One success story
already is SN~2011fe (PTF11kly), discovered only 11~hr after explosion
by PTF in M101, the Pinwheel Galaxy
\citep{Nugent11,Li11:ptf11kly}. The search for and
measurement of C in the many spectra of SN~2011fe obtained at extremely early
epochs will further our quest to better understand SNe~Ia.

\section*{Acknowledgments}

We thank R.~C.~Thomas for comments on earlier drafts of this
work. We are also grateful to the referee for suggestions that
improved the manuscript. Some of the data utilised herein were
obtained at the W. M. Keck  
Observatory, which is operated as a scientific partnership 
among the California Institute of Technology, the University of
California, and the National Aeronautics and Space Administration
(NASA); the observatory was made possible by the generous financial
support of the W. M. Keck Foundation. We wish to recognise
and acknowledge the very significant cultural role and reverence that
the summit of Mauna Kea has always had within the indigenous Hawaiian
community; we are most fortunate to have the opportunity to conduct
observations from this mountain.
We thank the staffs at the Lick and Keck Observatories for
their support during the acquisition of data. 
Financial support was received through U.S. NSF grant AST-0908886, DOE grants
DE-FC02-06ER41453 (SciDAC) and DE-FG02-08ER41563, and the TABASGO
Foundation. A.V.F. is grateful for the hospitality of the W. M. Keck
Observatory, where this paper was finalised.

\bibliographystyle{mn2e}

\bibliography{astro_refs}

\label{lastpage}

\end{document}